\documentclass[article]{emulateapj}
\usepackage{amsmath}

\slugcomment{submitted; \today}

\newcommand{\G}{\Gamma}

\newcommand{\sT}{\sigma_{\rm T}}
\newcommand{\p}{^\prime}
\newcommand{\e}{\epsilon}

\newcommand{\g}{\gamma}
\newcommand{\gp}{\gamma^{\prime}}

\newcommand{\ep}{\epsilon^\prime}

\newcommand{\dD}{\delta_{\rm D}}

\newcommand{\psim}{\lower.5ex\hbox{$\; \buildrel \propto \over\sim \;$}}
\newcommand{\lbar}{\lower.0ex\hbox{$\; \buildrel{\lower0.0ex \hbox{-}} \over\lambda  \;$}}

\newcommand{\tilr}{\tilde{r}}

\newcommand{\cm}{\mathrm{cm}}
\newcommand{\km}{\mathrm{km}}
\newcommand{\erg}{\mathrm{erg}}

\newcommand{\s}{\mathrm{s}}

\newcommand{\Hz}{\mathrm{Hz}}
\newcommand{\pc}{\mathrm{pc}}

\newcommand{\Mpc}{\mathrm{Mpc}}

\newcommand{\Gauss}{\mathrm{G}}

\newcommand{\fermi}{{\em Fermi}}

\shorttitle{The Properties of Parsec-Scale Blazar Jets}
\shortauthors{Finke}

\begin{document}
\title{The Properties of Parsec-Scale Blazar Jets}

\author{Justin D.\ Finke}

\affil{U.S.\ Naval Research Laboratory, Code 7653, 4555 Overlook Ave.\ SW,
        Washington, DC,
        20375-5352\\
}

\email{justin.finke@nrl.navy.mil}

\begin{abstract}

I show that by assuming a standard Blandford-K\"onigl jet, it is
possible to determine the bulk Lorentz factor and angle to the line of
sight of self-similar parsec-scale blazar jets by using five measured
quantities: redshift, core radio flux, extended radio flux, the
magnitude of the core shift between two frequencies, and apparent jet
opening angle.  From the bulk Lorentz factor and angle computed with
this method, one can compute other jet properties such as the Doppler
factor, magnetic field strength, and intrinsic jet opening angle.  I
use data taken from the literature and marginalize over nuisance
parameters associated with the electron distribution and equipartition
to compute these quantities, although the errors are large.  Results
are generally consistent with constraints from other methods.  Primary
sources of uncertainty are the errors on the core shift measurements
and the uncertainty in the electron spectral index.

\end{abstract}


\section{Introduction}
\label{intro}

Blazars are active galactic nuclei with relativistic jets oriented
close to our line of sight.  They are associated with compact jet
components on the milliarcsecond scale that can be resolved with radio
very long baseline interferometry (VLBI).  VLBI images of these
objects reveal a stationary core from which knots emerge.  Often
the apparent speeds of these knots projected on the sky
$\beta_{\rm app}c>1$, i.e., they appear to be moving faster than the
speed of light $c$.  This is a well-known optical illusion caused by
motion with intrinsic speed $\beta c$ close to $c$ and angle to
the line of sight $\theta \ll 1$ \citep{rees66}.

One of the defining features of blazars is their bright, stationary
cores seen in VLBI images.  This core is generally thought to be
described by the Blandford-K\"onigl (BK) model
\citep{blandford79,konigl81}.  In this model, the core emission is
the superposition of self-absorbed components in a
steady, continuous parsec-scale jet.  This model is likely a useful
approximation to reality: a model where a number of colliding shells
in the jet accelerate electrons, which cool through radiative and
adiabatic losses can reproduce many of the features of the BK model
\citep{jamil10}.

The BK model makes two key predictions.  The first is flat radio
spectra ($\alpha\approx0$, where the radio flux density $F_\nu \propto
\nu^{-\alpha}$ and $\nu$ is the observed frequency). The observation
of flat spectra in blazar cores was the primary empirical motivation
for the model.  The second is a frequency-dependent core position;
that is, the core's position on the sky will ``shift'' between two
different positions when viewed at different frequencies.  This effect
has also been observed in a number of blazar radio cores
\citep[e.g.,][]{lobanov98,kovalev08,osullivan09,sokolovsky11,pushkarev12}.
When observed at multiple frequencies, the magnitude of the core shift
is observed to be $\propto \nu^{-1}$, in agreement with the BK model
prediction \citep{osullivan09,sokolovsky11}.

The magnitudes of the core shifts have been used to infer the magnetic
field of the jet \citep[e.g.,][]{pushkarev12,zdz15}.  This usually
requires making some assumption about the jet speed (bulk Lorentz
factor, $\G$) or orientation (angle to the line of sight, $\theta$).
Here I show that it is in fact not necessary to make assumptions about
the speed and orientation of blazar jets.  Assuming the BK model is a
reasonable description of these jets, one can determine $\G$,
$\theta$, and other jet parameters such as the magnetic field, from
five observables: the redshift $z$, the core flux density $F_\nu$, the
core shift $\Delta\phi$, the apparent opening angle for the jet
$\alpha_{\rm app}$, and the extended radio flux, which is used as a
proxy for jet power \citep[e.g.,][]{birzan04,birzan08,cavagnolo10}.

In Section \ref{model} I describe the BK jet model and show how it can
be used to determine jet parameters from the observables.  In Section
\ref{results} this model is applied to a sample of blazar radio jets
with appropriate observations from the literature.  The measured
properties of their jets, including $\G$ and $\theta$ are presented.
In Section \ref{implications} I compare my results with previous
estimates of these parameters, and explore some of the implications of
the results.  Finally, I conclude with a discussion in Section
\ref{discussion}.

\section{Continuous Jet Model}
\label{model}

\subsection{Synchrotron Self-Absorption}

In the $\delta$-function approximation, synchrotron self-absorption
(SSA) opacity at a dimensionless energy $\e=h\nu/(m_ec^2)\approx
\nu/(1.23\times10^{20}\ \Hz)$ is
\begin{flalign}
\label{kappa1}
\kappa(\e) = \frac{-\pi}{36} \frac{\lambda_C r_e}{\e} 
\left\{ \g \frac{ \partial}{\partial\g}
\left[ \frac{n_e(\g)}{\g^2}\right]\right\}
\Biggr|_{\g=\sqrt{\e/(2\e_B)}}\ 
\end{flalign}
\citep{dermer09_book} where $r_e\approx2.8\times10^{-13}\ \cm$ is the
classic electron radius, $\lambda_C\approx2.4\times10^{-10}\ \cm$,
$\e_B=B/B_c$, $B$ is the magnetic field, and
$B_c=4.414\times10^{13}\ \Gauss$.  Let the electron density
distribution be a power-law, $n_e(\g)=n_0 \g^{-p} H(\g; \g_1, \g_2)$
where
\begin{equation}
H(x; a, b) = \left\{ \begin{array}{ll}
1 & a < x <b \\
 0 & \mathrm{otherwise}
\end{array}
\right. \ .
\end{equation}
Then
\begin{flalign}
\label{kappa2}
\kappa(\e) & = C(p) \lambda_c r_e \e^{-(p+4)/2} \e_B^{(p+2)/2} n_0\ 
\\ \nonumber &\times
H[\sqrt{\e/(2\e_B)}; \g_1, \g_2]
\end{flalign}
where
\begin{flalign}
C(p) = \frac{\pi}{18} (p+2)(2)^{p/2}\ .
\end{flalign}
This can be compared with more precise expressions by, e.g.,
\citet{gould79} and \citet{zdz12_tail}.

\subsection{Continuous Jet}

\begin{figure*}
\vspace{10.0mm} 
\epsscale{1.0} 
\plotone{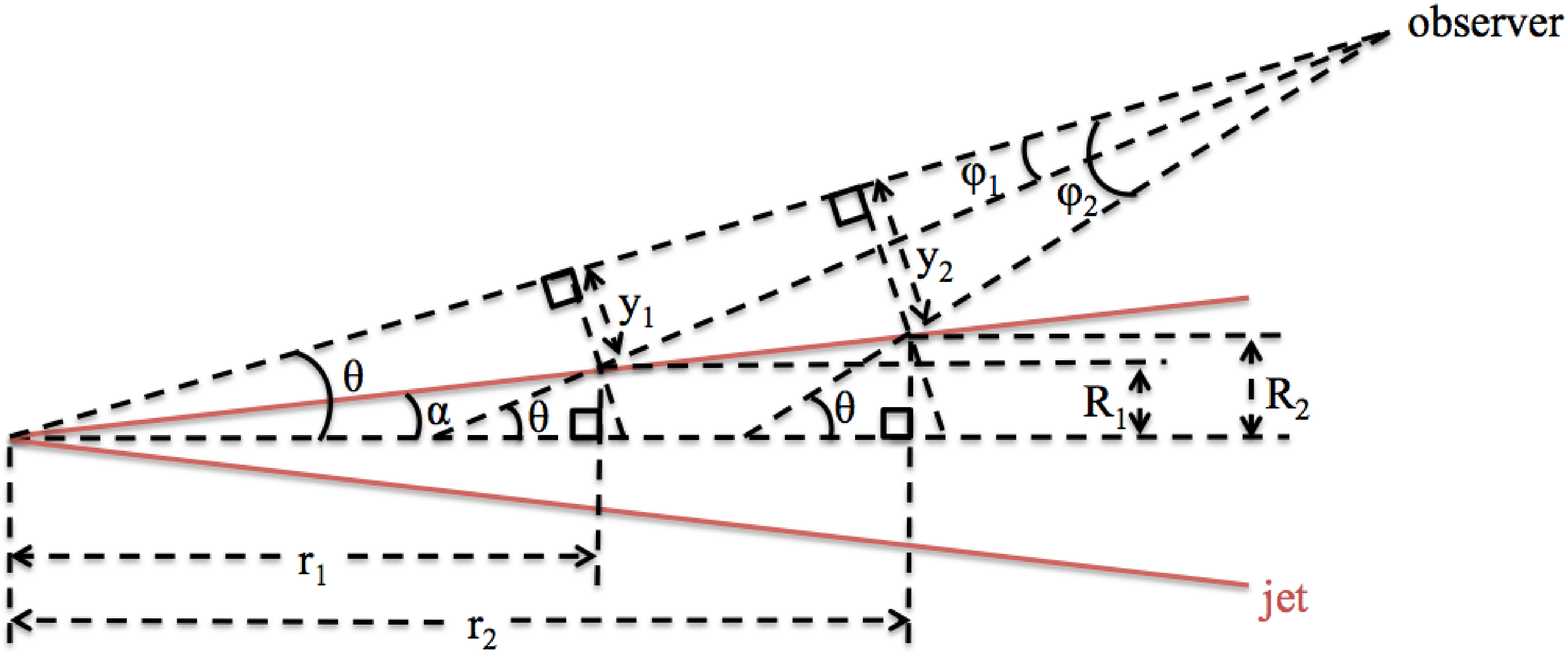}
\caption{Cross-section illustration of conical jet geometry.}
\label{geometryfig}
\vspace{2.2mm}
\end{figure*}

Consider a conical continuous relativistic jet with half-opening angle
$\alpha \ll 1$; see Figure \ref{geometryfig} for an illustration of
the geometry.  The source has cosmological redshift $z$ giving it a
luminosity distance
\begin{flalign}
d_L = (1+z) \frac{c}{H_0} \int_0^z 
\frac{dz\p}{\sqrt{ \Omega_m(1+z\p)^3 + \Omega_\Lambda} }\ .
\end{flalign}
I use $H_0=70\ \km\ \s^{-1}\ \Mpc^{-1}$, $\Omega_m=0.3$, and
$\Omega_\Lambda=0.7$. If a core shift is observed it implies $\alpha <
\theta$.  Since I am interested in the case where a core shift is
indeed observed, I assume this is the case.  The distance from the
base of the jet is $r$ and the half width of the jet at $r$ is $R$, so
that $R/r = \tan\alpha \approx \alpha$.  The jet moves with speed
$\beta c$ giving it a bulk Lorentz factor $\G = (1-\beta)^{-1/2}$.
The electron density and magnetic field in the jet are assumed to
decrease with $r$, so that
\begin{flalign}
n_0 = n_{00}\ \tilr^{-a}\ ,
\\
\e_B = \e_{B0}\ \tilr^{-b}
\end{flalign}
where $\tilr = r/r_0$, $r_0$ is some reference distance along the
jet, $n_{00} = n_0(r_0)$, and $\e_{B0} = \e_B(r_0)$.  The half-jet width
at $r_0$ is $R_0=\alpha r_0$.

An observer sees the jet at an angle $\theta$ to the center of the jet
axis, so that the Doppler factor $\dD=[\G(1-\beta\cos\theta)]^{-1}$
and the apparent half-opening angle 
\begin{flalign}
\label{alphaapp}
\alpha_{\rm  app} = \alpha/\sin\theta\ .  
\end{flalign}
Hereafter, in the frame co-moving with the jet,
  quantities are primed, such as the co-moving distance from the jet
  base, $r\p$.  Quantities in the rest frame of the galaxy are
  unprimed, except for energies, where unprimed quantities are in the
  observer frame, so that $\e=\ep\dD/(1+z)$.  The difference between
  the observer frame and galaxy frame is a factor $1+z$. 

\subsection{Synchrotron Self-Absorption in Continuous Jet}

The absorption optical depth
\begin{flalign}
\label{tau1}
\tau(\ep) = \int d\ell\p\ \kappa\p(\ep)
\end{flalign}
where $\ell\p$ is the distance a photon travels in the jet to the
observer, in the co-moving frame.  The angle to the
  line of sight in the comoving frame is transformed as $\sin\theta\p
  = \dD\sin\theta$ \citep[e.g.,][]{rybicki79,zdz12_tail}.  In the frame of
  the galaxy $\ell = R/\sin\theta = r\alpha_{\rm app}$.
Using Equation (\ref{kappa2}) for $\kappa(\ep)$, the SSA opacity, and
with a change of variables from $\ell\p$ to $r$,
Equation (\ref{tau1}) can be integrated to give 
\begin{flalign}
\label{tau2}
\tau(\e) & = C(p) \alpha_{\rm app} k_1^{-1} \lambda_C r_e (1+z)^{-(p+4)/2} \dD^{(p+2)/2}
\\ \nonumber & \times
\e_{B0}^{(p+2)/2}\e^{-(p+4)/2} \tilr^{-k_1} n_{00} r_0
\end{flalign}
where $k_1 = b(p+2)/2 + a - 1$.  Equation (\ref{tau2}) can
be compared to Equation (1) of \citet{lobanov98} and Equation (26) of
\citet{blandford79}.  By setting $\tau(\e)=1$ in Equation (\ref{tau2})
one can solve for the energy where the jet becomes optically thin to
synchrotron self-absorption,
\begin{flalign}
\label{essa}
\e_{\rm SSA} = \e_{\rm SSA,0}\ \tilr^{-k_2}
\end{flalign}
where
\begin{flalign}
\label{essa0}
\e_{\rm SSA,0} = &
\left[ \frac{C(p)\lambda_C r_e n_{00} R_0}{\sin\theta k_1} \right]^{2/(p+4)}
\\ \nonumber & \times
(\dD\e_{B0})^{(p+2)/(p+4)} \frac{1}{1+z}\ ,
\end{flalign}
\begin{flalign}
k_2 = \frac{2k_1}{p+4} = \frac{ 2a + b(p+2) - 2}{p+4}\ ,
\end{flalign}
and $R_0 = \alpha r_0$.

\subsection{Core Shift}

In the context of the BK model, the core is the surface where
$\tau=1$.  The observed angular distance of a core from the base of the jet 
observed at a given observed dimensionless energy $\e_{\rm SSA}$ is 
\begin{flalign}
\phi = \frac{r \sin\theta}{d_A} = \frac{R_0}{\alpha_{\rm app} d_A} 
\left( \frac{\e_{SSA,0}}{\e_{SSA}}\right)^{1/k_2}
\end{flalign}
where $d_A = d_L/(1+z)^2$ is the angular diameter distance.  The core
shift between two dimensionless energies $\e_{\rm SSA,1}$ and $\e_{\rm
  SSA,2}$ where $\e_{\rm SSA,2} < \e_{\rm SSA,1}$ is then
\begin{flalign}
\label{dphi1}
\Delta\phi & = \phi_2 - \phi_1 = \frac{R_0(1+z)^2}{\alpha_{\rm app} d_L} \e_{SSA,0}^{1/k_2}
\left( \frac{1}{\e_{SSA,2}^{1/k_2}} - \frac{1}{\e_{SSA,1}^{1/k_2}} \right)
\nonumber \\ 
& = \frac{R_0(1+z)^2}{\alpha_{\rm app} d_L}
\left( \frac{\e_{\rm SSA,0}}{\e_{\rm eff}}\right)^{1/k_2}
\end{flalign}
where
\begin{flalign}
\e_{\rm eff} = \frac{ \e_{\rm SSA,2} \e_{\rm SSA,1}}
{ \left( \e_{\rm SSA,1}^{1/k_2} - \e_{\rm SSA,2}^{1/k_2}\right)^{k_2} }\ .
\end{flalign}
The geometry of this core shift is illustrated in Figure
\ref{geometryfig}.  For the ``standard'' continuous BK jet, $a=2$ and
$b=1$ \citep{blandford79,konigl81} gives $k_2=1$, in agreement with
core shift observations \citep{osullivan09,sokolovsky11}.

\subsection{Flux of Continuous Jet}

For a continuous relativistic jet with stationary
  pattern, the observed $\nu F_\nu$ flux
\begin{flalign}
\label{fesy1}
f_\e^{sy} = \frac{\dD^3}{4\pi d_L^2\G} \int dV\p\ \ep\ j\p(\ep) \ ,
\end{flalign}
\citep{sikora97,zdz12_tail,zdz12_gamma} where $\ep j\p(\ep)$ is the comoving frame emissivity
(from synchrotron in this case), and the comoving volume $dV\p = \G
dV$.  For a conical jet, $dV = \pi R^2 dr$.  It is well-known that for
synchrotron emission, $F_\nu \propto \nu^{5/2}$ in the SSA optically
thick ($\tau > 1$) regime and $F_\nu \propto \nu^{-(p-1)/2}$ in the SSA
optically thin ($\tau < 1$) regime.  To account for both these
regimes, I approximate the emission from a jet volume element $dV\p$
as
\begin{eqnarray}
\label{eje}
 \ep j\p(\ep) \approx 2 c \sT u_B / 3
\nonumber \\  \times
\left\{ \begin{array}{ll}
\g^{\prime 3}_{\rm SSA} n_e(\gp_{\rm SSA}) \left(\ep/\ep_{\rm SSA}\right)^{7/2} & \ep < \ep_{\rm SSA} \\
\g^{\prime 3} n_e(\gp) & \ep_{\rm SSA} < \ep \\
\end{array}
\right. \ ,
\end{eqnarray}
where 
\begin{flalign}
\gp = \sqrt{\frac{\ep}{\e_B}}\ ,
\end{flalign}
\begin{flalign}
\gp_{\rm SSA} = \sqrt{\frac{\ep_{\rm SSA}}{\e_B}}\ ,
\end{flalign}
\begin{flalign}
u_B=\frac{B^2}{8\pi} = u_{\rm Bcr} \e_B^2
\end{flalign}
is the Poynting flux energy density, and
\begin{flalign}
u_{\rm Bcr}=\frac{B_c^2}{8\pi} \ .
\end{flalign}
Note that $B$ and $\e_B$ are in the comoving frame, although I
neglect the primes on them.  Substituting Equations (\ref{eje}) into
Equation (\ref{fesy1}),
\begin{flalign}
f_\e^{sy} & = \frac{\dD^3 n_0 c \sT u_{\rm Bcr}R^2}{6 d_L^2}
\nonumber \\ & \times
\Biggr\{ \int_{r_{\min}}^{r_b} dr\ \e_B^{(1+p)/2} \e_{\rm SSA}^{\prime (3-p)/2}
\left(\frac{\ep}{\ep_{\rm SSA}}\right)^{7/2} 
\nonumber \\ 
& + \int_{r_b}^{r_{\max}} dr\ \e_B^{(1+p)/2}\e^{\prime (3-p)/2} \Biggr\}\ ,
\end{flalign}
where $r_{\min}$ is the distance from the base of the jet where
emission begins, $r_{\max}$ is the distance from the base of the jet
where emission ends, and $r_b = (\ep_{\rm SSA,0}/\ep)^{1/k_2}$.  Using
Equation (\ref{essa}) and performing the integrals assuming $r_{\min}
\ll r_b \ll r_{\max}$,
\begin{flalign}
\label{fsy1}
f_\e^{sy} & = \frac{\dD^3 c \sT u_{\rm Bcr} R_0^2 r_0 n_{00} \e_{B0}^{(1+p)/2}}{3 d_L^2}
\nonumber \\ & \times
\Biggr\{ \frac{\e^{\prime 7/2} \e_{\rm SSA,0}^{\prime -(p+4)/2} }{6 + k_2(p+4) - b(p+1) - 2a}
\nonumber \\ & \times
\left( \frac{\ep_{SSA,0}}{\ep}\right)^{\frac{6+k_2(p+4)-b(p+1)-2a}{2}} 
\nonumber \\ &
+ \frac{\e^{\prime (3-p)/2}}{b(p+1)+2a-6} 
\left( \frac{\ep_{\rm SSA,0}}{\ep}\right)^{\frac{6-b(p+1)-2a}{2}} \Biggr\}\ .
\end{flalign}
For the standard BK jet, $a=2$, $b=1$, and $k_2=1$, 
\begin{flalign}
\label{fsy2}
\frac{f_\e^{sy}}{\e} & = \dD^{(3+p)/2} c\sT u_{\rm Bcr} R_0^2 r_0 n_{00} \e_{B0}^{(1+p)/2}\e_{SSA,0}^{(1-p)/2}
\nonumber \\ & \times
\left[\frac{ (1+z)^{(3-p)/2}}{3d_L^2} \right]
\left[ \frac{p+4}{5(p-1)} \right]\ .
\end{flalign}
The flux density $F_\nu$ can be found directly from Equation (\ref{fsy2}), since
\begin{flalign}
F_\nu = \frac{f_\e^{sy}}{\e} \frac{h}{m_e c^2}\ .
\end{flalign}
Thus, $F_\nu$ will be independent of $\nu$ (or $\e$), i.e., ``flat''
in agreement with observations of the radio spectra of the cores of
blazars.  Equation (\ref{fsy2}) agrees with similar formulations by a
number of other authors
\citep[e.g.,][]{blandford79,konigl81,falcke95,zdz12_tail}.  This
  calculation only accounts for a single jet traveling towards the
  observer, neglecting the counter-jet emission
  \citep[cf.][]{zdz12_tail}.

\subsection{Jet Power}
\label{jetpowersection}

The jet power from a single jet
\begin{flalign}
\label{Pj}
P_j & = \hat\g_{\rm ad} \pi R^2 \G^2 \beta c 
\nonumber \\ & \times
\left(g_B u_B + u_e + u_p + 
\frac{\G-1}{\hat\g_{\rm ad}\G}\rho c^2\right)
\end{flalign}
where $\hat\g_{\rm ad}$ is the adiabatic index, $g_B\sim 1$ is a factor taking
into account the geometry of the magnetic field,
\begin{flalign}
u_e = m_e c^2 \int d\g\ \g\ n_e(\g) = m_ec^2 n_0 A(p, \g_1, \g_2)\ 
\end{flalign}
is the electron energy density, 
\begin{equation}
A(p, \g_1, \g_2) = \left\{ \begin{array}{ll}
(p-2)^{-1}(\g_1^{2-p} - \g_2^{2-p}) & p \ne 2 \\
\ln(\g_2/\g_1) & p = 2 
\end{array}
\right. \ ,
\end{equation}
$u_p$ is the relativistic proton energy density, 
  and $\rho$ is the mass density \citep[see][for a description of this
    term]{bicknell94,zdz14}.  The adiabatic index in
Equation (\ref{Pj}) takes into account the contribution from both the
pressure and energy density
\citep[e.g.][]{levinson06,zdz12_tail,zdz14}.  I use $\hat\g_{\rm
  ad}=4/3$ for a relativistic plasma and $g_B=1.5$
  for a toroidal magnetic field
  \citep[e.g.,][]{levinson06,zdz15}.  I define 
  $\xi_e \equiv u_e/u_B$, $\xi_p \equiv u_p/u_B$, and
  $\xi_m \equiv \rho c^2/(\hat\g_{\rm ad} u_B)$, so that
\begin{eqnarray}  
P_j & = \hat\g_{\rm ad} \pi R^2 \G^2 \beta c u_B
\nonumber \\ & \times
\left(g_B + \xi_e + \xi_p + \frac{\G-1}{\G}\xi_m \right) \ .
\end{eqnarray}

The jet power $P_j$ can be estimated from the power needed to inflate
a cavity in the hot X-ray emitting intracluster medium, and is
correlated with the extended radio luminosity of a radio-loud AGN
\citep[e.g.,][]{birzan04,birzan08,cavagnolo10}.  The relationship
between the jet power $P_j$ and the 200-400 MHz extended luminosity
$L_{\rm ext}$ is 
\begin{flalign}
\label{powercavag}
\log_{10}\left[\frac{P_j}{\erg\ \s^{-1}}\right] = 
c_1 \left\{\log_{10}\left[\frac{L_{\rm ext}}{\erg\ \s^{-1}}\right]-40\right\} +
c_2\ ,
\end{flalign}
where $c_1=0.64\pm0.09$ and $c_2=43.54\pm0.12$
\citep{cavagnolo10}.  I use this expression to estimate the jet power
from $L_{\rm ext}$.  I divide the power obtained from Equation
  (\ref{powercavag}) by 2 to account for only a single jet.

Using Equation (\ref{Pj}), 
Equation (\ref{essa0}) can be written 
\begin{flalign}
\label{essa0_2}
\e_{SSA,0} & = \left[ \frac{C(p) \lambda_C r_e }
{\sin\theta k_1 m_ec^2 A(p, \g_1, \g_2)\G^2} \right]^{\frac{2}{p+4}}
\nonumber \\ & \times
\frac{1}{1+z} \left(\frac{\dD}{\G}\right)^{\frac{p+2}{p+4}} 
\left(\frac{P_j}{\hat\g_{\rm ad} \pi \beta c}\right)^{\frac{p+6}{2(p+4)}}
\nonumber \\ & \times
\frac{1}{R_0\chi_1}\left(\frac{1}{u_{\rm Bcr}}\right)^{\frac{p+2}{2(p+4)}}\ ,
\end{flalign}
where 
\begin{flalign}
\chi_1 & = \xi_e^{\frac{-2}{p+4}}
\left(g_B+\xi_e+\xi_p+\xi_m\frac{\G-1}{\G}\right)^{\frac{p+6}{2(p+4)}}\ .
\end{flalign}
For the $a=2$, $b=1$ model, 
\begin{flalign}
\label{Pj2}
P_j & = \hat\g_{\rm ad} \pi R_0^2\G^2\beta c u_{\rm Bcr} \e_{B0}^2
\nonumber \\ & \times
\left(g_B+\xi_e+\xi_p\xi_m\frac{\G-1}{\G}\right)
\nonumber \\ & 
= \hat\g_{\rm ad} \pi R_0^2\G^2\beta c m_ec^2 n_{00} A(p, \g_1, \g_2)
\nonumber \\ & \times
\left(1+\frac{g_B}{\xi_e} + \frac{\xi_p}{\xi_e} + 
\frac{\xi_m}{\xi_e}\frac{\G-1}{\G}\right)\ .
\end{flalign} 

\subsection{Determination of Jet Parameters}
\label{determinejet}

\begin{figure}
\vspace{10.0mm} 
\epsscale{1.0} 
\plotone{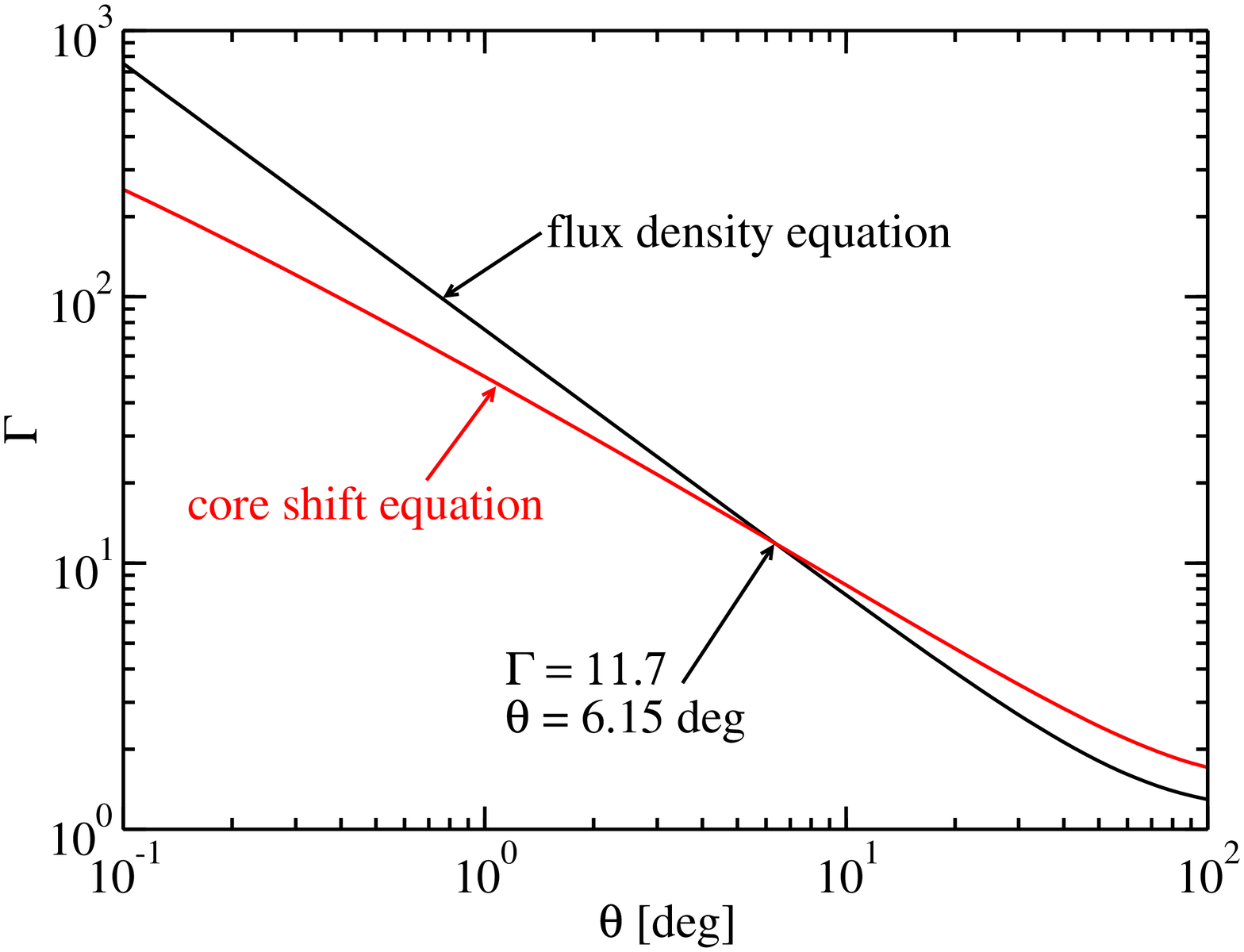}
\caption{A plot of $\G$ versus $\theta$ for a particular monte
    carlo iteration for 2251+158 (3C 454.3) using the core shift
  equation (Equation [\ref{dphi2}]) and the flux density equation
  (Equation [\ref{fsy3}]) with the observed parameters given in Table
  \ref{table_data}.  The intersection of these curves gives the
  numerical solution; in this case, $\G=11.7$ and
  $\theta=6.15\arcdeg$.}
\label{solnplot}
\vspace{2.2mm}
\end{figure}

Combining Equations (\ref{dphi1}) and (\ref{essa0_2}) for the the $a=2$, $b=1$ model, with
some algebraic manipulation, 
\begin{flalign}
\label{dphi2}
\Delta\phi & = \frac{1+z}{\e_{\rm eff}\alpha_{\rm app} d_L \chi_1} 
\left[ \frac{C(p) \lambda_C r_e}{\sin\theta k_1 m_ec^2 A(p, \g_1, \g_2) \G^2} \right]
^{\frac{2}{p+4}}
\nonumber \\ &\times
\left[ \frac{\dD}{\G}\right]^{\frac{p+2}{p+4}}
\left[ \frac{P_j}{\hat\g_{\rm ad}\pi\beta c}\right]^{\frac{p+6}{2(p+4)}}
\left[ \frac{1}{u_{\rm Bcr}}\right]^{\frac{p+2}{2(p+4)}}\ .
\end{flalign}
Similarly, combining Equations (\ref{fsy2}), 
  (\ref{dphi1}), and (\ref{Pj2}), and more algebraic manipulation,
\begin{flalign}
\label{fsy3}
\frac{f_\e^{sy}}{\e} & = \frac{c\sT}{3\chi_2 A(p, \g_1, \g_2) m_e c^2\dD\sin\theta } 
\frac{p+4}{5(p-1)} u_{\rm Bcr}^{\frac{3-p}{4}}
\nonumber \\ & \times
\left[ \frac{1+z}{\alpha_{\rm app}}\right]^{\frac{1+p}{2}}
\left[ \left(\frac{\dD}{\G}\right)^2 \frac{P_j}{\hat\g_{\rm ad}\pi\beta c}\right]^{\frac{5+p}{4}}
\nonumber \\ &\times
\left[ \Delta\phi \e_{\rm eff}\right]^{\frac{1-p}{2}}
d_L^{\frac{-(3+p)}{2}}
\end{flalign}
where
\begin{flalign}
\chi_2 & = \frac{1}{\xi_e}
\left(g_B + \xi_e + \xi_p + \xi_m\frac{\G-1}{\G}\right)^{\frac{5+p}{4}}\ .
\end{flalign}
Equations (\ref{dphi2}) and (\ref{fsy3}) have two physical unknown
parameters: $\G$ and $\theta$; and five observables: $z$, $F_\nu$,
$\Delta\phi$, $\alpha_{\rm app}$, and $P_j$ (through $L_{\rm ext}$).
These two equations can be solved numerically for $\G$ and $\theta$.
I have several ``nuisance parameters'' to marginalize over: $p$,
$\g_1$, $\g_2$, $\xi_e$, $\xi_p$.  An example of this calculation is
given in Figure \ref{solnplot}.  Here $\G$ versus $\theta$ are plotted
from Equations (\ref{dphi2}) and (\ref{fsy3}) using the observations
given in Table \ref{table_data} for single randomly
  drawn values of the nuisance parameters.  The
intersection of the curves is the numerical solution, giving $\G$ and
$\theta$.

Once $\G$ and $\theta$ are known, other parameters can be computed as
well.  For instance, $\dD=[\G(1-\beta\cos\theta)]^{-1}$; the
intrinsic half-opening angle, $\alpha$ can be computed from Equation
(\ref{alphaapp}); and $B_0$ can be computed from Equation (\ref{Pj2})
for a given $r_0$ (recall $B_0=\e_{B0}B_c$).  The apparent jet speed
an observed would see from the flow can also be computed,
\begin{flalign}
\label{betaappeqn}
\beta_{\rm app} = \sqrt{ 2\dD\G - \dD^2 -1}\ .
\end{flalign}

\section{Results}
\label{results}

\subsection{Data}

The measurement data taken from the literature can be found in Table
\ref{table_data}.  The most prominent VLBI instrument is the Very Long
Baseline Array (VLBA), spread throughout North America.  In general, I
rely heavily on data taken as part of the Monitoring of Jets of Active
Galactic Nuclei with VLBA Experiments (MOJAVE).  I thus use the MOJAVE
collaboration's optical spectral classification for sources (BL Lac
object, flat spectrum radio quasar [FSRQ], or Narrow Line Seyfert 1)
and redshifts.  For a discussion of redshifts and their measurements,
particularly BL Lacs that often have unknown or poorly known
redshifts, see \citet{lister11}.

The core shift measurements $\Delta\phi$ from the MOJAVE program
between 15 and 8 GHz were taken from \citet{pushkarev12}.  Their
errors were taken to be 51 $\mu$as, as discussed by those authors.
The core flux densities $F_\nu$ at $\nu=15$\ GHz were taken from the
MOJAVE website\footnote{\url{http://www.physics.purdue.edu/MOJAVE/}}
\citep[e.g.,][]{lister09}.  The flux densities used were the ones
measured during the same epoch the core shift measurement was
performed.  I assumed there was no error on $F_\nu$, since the errors
on these values are likely small compared to errors on other
observables.  Apparent jet opening angles $\alpha_{\rm app}$, measured
as part of the MOJAVE program, were taken from \citet{pushkarev09}.
See also \citet{pushkarev17}.  The errors were assumed to be 10\%,
based on the different results between how the opening angles were
measured, in the image plane or in the $(u,v)$ plane, as described by
\citet{pushkarev09}.  The extended radio luminosities, $L_{\rm ext}$,
are taken from a compilation by \citet{meyer11}.  They found $L_{\rm
  ext}$ from a spectral decomposition method, separating the core and
extended luminosity using spectral energy distribution (SED) shapes.
The errors on $L_{\rm ext}$ were also assumed to be negligible.

I found there were 64 sources that were found to have data from all of
these sources, including 11 BL Lac objects, 52 FSRQs, and 1 Narrow
Line Seyfert 1.

\begin{deluxetable*}{llcccccc}
\tabletypesize{\scriptsize}
\tablecaption{ Blazar Radio Measurements}
\tablewidth{0pt}
\tablehead{
\colhead{Source} &
\colhead{Alias} &
\colhead{Type\tablenotemark{a}} &
\colhead{$z$} &
\colhead{$\log_{10}\left[\frac{L_{\rm ext}}{\erg\ \s^{-1}}\right]$} &
\colhead{$F_\nu({\rm core})$ [Jy]} &
\colhead{$2\alpha_{\rm app}$ [\arcdeg]} &
\colhead{$\Delta\phi$ [mas]} 
}
\startdata
0133$+$476&  DA 55         &   Q&    0.859&	41.93&	1.781&	21.7&	 	0.099	\\
0202$+$149&  4C +15.05     &   Q&	0.405&	41.39&	0.921&	16.4&	 	0.113	\\
0212$+$735&  S5 0212$+$73  &   Q&	2.367&	42.30&	3.281&	16.4&	 	0.143	\\
0215$+$015&  OD 026        &   Q&	1.715&	43.52&	1.170&	36.7&	 	0.111	\\
0234$+$285&  4C 28.07      &   Q&	1.207&	43.21&	2.944&	19.8&	 	0.239	\\
0333$+$321&  NRAO 140      &   Q&	1.259&	42.98&	1.343&	8.0 &	 	0.276	\\
0336$-$019&  4C 28.07      &   Q&	0.852&	42.36&	2.311&	26.8&	 	0.105	\\
0403$-$132&  PKS 0403$-$13 &   Q&	0.571&	43.25&	1.808&	16.4&	 	0.285	\\
0420$-$014&  PKS 0420$-$01 &   Q&	0.914&	42.66&	3.746&	22.7&	 	0.256	\\
0528$+$134&  PKS 0528$+$134&   Q&	2.070&	43.73&	3.847&	16.1&	 	0.150	\\
0605$-$085&  OC $-$010     &   Q&	0.872&	42.94&	1.120&	14.0&	 	0.096	\\
0607$-$157&  PKS 0607$-$15 &   Q&	0.324&	39.84&	3.983&	35.1&	 	0.254	\\
0716$+$714&  S5 0716$+$71  &   B&	0.310&	41.98&	0.586&	17.2&	 	0.125	\\
0738$+$313&  OI 363        &   Q&	0.631&	42.17&	1.226&	10.5&	 	0.138	\\
0748$+$126&  OI 280	   &   Q&    0.889&	42.77&	3.821&	16.2&	 	0.097	\\
0754$+$100&  PKS 0754$+$100&   B&	0.266&	41.44&	1.304&	13.7&	 	0.280	\\
0804$+$499&  OJ 508        &   Q&	1.436&	42.34&	0.458&	35.3&	 	0.073	\\
0805$-$077&  PKS 0805$-$07 &   Q&	1.837&	43.78&	1.08 &  18.8&	 	0.228	\\
0823$+$033&  PKS 0823$+$033&   B&	0.506&	40.95&	1.828&	13.4&	 	0.142	\\
0827$+$243&  OJ 248        &   Q&	0.940&	42.62&	0.775&	14.6&	 	0.139	\\
0829$+$046&  OJ 049        &   B&	0.174&	41.00&	0.576&	18.7&	 	0.131	\\
0836$+$710&  4C +71.07     &   Q&	2.218&	43.91&	2.222&	12.4&	 	0.172	\\
0906$+$015&  4C +01.24     &   Q&    1.024&	42.67&	1.846&	17.5&	 	0.203	\\
0917$+$624&  OK 630        &   Q&    1.446&	43.07&	1.017&	15.9&	 	0.111	\\
0945$+$408&  4C +40.24     &   Q&	1.249&	43.64&	0.948&	14.0&	 	0.113	\\
1038$+$064&  4C +06.41     &   Q&	1.265&	42.86&	1.211&	6.7 &	 	0.146	\\
1045$-$188&  PKS 1045$-$18 &   Q&	0.595&	43.19&	1.172&	8.0 &	 	0.167	\\
1101$+$384&  Mrk 421       &   B&	0.031&	39.59&	0.343&	18.2&	 	0.254	\\
1127$-$145&  PKS 1127$-$14 &   Q&	1.184&	43.24&	3.308&	16.1&	 	0.089	\\
1150$+$812&  S5 1150$+$81  &   Q&	1.250&	43.44&	1.710&	15.0&	 	0.082	\\
1156$+$295&  4C +29.45     &   Q&	0.729&	42.95&	3.596&	16.7&	 	0.154	\\
1219$+$044&  4C +04.42     &   N&	0.965&	42.94&	1.178&	13.0&	 	0.169	\\
1222$+$216&  4C +21.35     &   Q&	0.432&	42.88&	0.507&	10.8&	 	0.170	\\
1308$+$326&  OP 313        &   Q&	0.997&	42.86&	0.917&	18.5&	 	0.095	\\
1334$-$127&  PKS 1335$-$127&   Q&	0.539&	42.51&	4.714&	12.6&	 	0.274	\\
1413$+$135&  PKS B1413$+$135&   B&	0.247&	40.32&	0.731&	8.8 &	 	0.228	\\
1502$+$106&  OR 103	   &   Q&    1.839&	43.30&	1.510&	37.9&	 	0.056	\\
1504$-$166&  PKS 1504$-$167&   Q&	0.876&	42.41&	1.162&	18.4&	 	0.115	\\
1510$-$089&  PKS 1510$-$08 &   Q&	0.360&	42.17&	1.718&	15.2&	 	0.151	\\
1538$+$149&  4C +14.60     &   B&	0.605&	42.72&	1.033&	16.1&	 	0.077	\\
1606$+$106&  4C +10.45     &   Q&	1.226&	42.73&	1.462&	24.0&	 	0.073	\\
1611$+$343&  DA 406        &   Q&	1.397&	42.76&	4.892&	26.9&	 	0.059	\\
1633$+$382&  4C +38.41     &   Q&	1.814&	43.24&	2.419&	22.6&	 	0.139	\\
1637$+$574&  OS 562        &   Q&	0.751&	42.67&	1.413&	10.7&	 	0.103	\\
1641$+$399&  3C 345        &   Q&	0.593&	43.16&	5.279&	12.9&	 	0.201	\\
1652$+$398&  Mrk 501       &   B&	0.033&	39.46&	0.877&	19.5&	 	0.279	\\
1730$-$130&  NRAO 530      &   Q&	0.902&	43.40&	2.582&	10.4&	 	0.195	\\
1749$+$096&  4C +09.57     &   B&	0.322&	41.39&	4.585&	16.8&	 	0.083	\\
1807$+$698&  3C 371        &   B&	0.051&	40.59&	1.137&	11.0&	 	0.216	\\
1928$+$738&  4C +73.18     &   Q&	0.302&	42.28&	3.396&	9.8 &	 	0.155	\\
1936$-$155&  PKS 1936$-$15 &   Q&	1.657&	42.66&	0.691&	35.2&	 	0.236	\\
2121$+$053&  PKS 2121$+$053&   Q&	1.941&	42.46&	1.955&	34.0&	 	0.148	\\
2128$-$123&  PKS 2128$-$12 &   Q&	0.501&	41.90&	2.665&	5.0 &	 	0.242	\\
2131$-$021&  4C $−$02.81   &   Q&	1.285&	43.33&	2.005&	18.4&	 	0.099	\\
2134$+$004&  PKS 2134+004  &   Q&	1.932&	43.51&	6.198&	15.2&	 	0.172	\\
2155$-$152&  PKS 2155$-$152&   Q&	0.672&	42.82&	2.151&	17.6&	 	0.343	\\
2200$+$420&  BL Lac        &   B&	0.069&	39.86&	3.0  &  26.2&	 	0.052	\\
2201$+$171&  PKS 2201$+$171&   Q&	1.076&	43.05&	1.349&	13.6&	 	0.369	\\
2201$+$315&  4C +31.63     &   Q&	0.295&	42.16&	2.334&	12.8&	 	0.345	\\
2223$-$052&  3C 446        &   Q&	1.404&	44.11&	5.270&	11.7&	 	0.162	\\
2227$-$088&  PHL 5225      &   Q&	1.560&	42.55&	1.515&	15.8&	 	0.193	\\
2230$+$114&  CTA 102       &   Q&	1.037&	43.40&	2.268&	13.3&	 	0.320	\\
2251$+$158&  3C 454.3      &   Q&	0.859&	43.62&	12.541&	40.9&	 	0.159	\\
2345$-$167&  PKS 2345$-$16 &   Q&	0.576&	42.61&	2.280&	15.8&	 	0.157	
\enddata
\tablenotetext{a}{Optical Spectral Type.  B = BL Lac object, Q = FSRQ,
  N = Narrow Line Seyfert 1.}
\label{table_data}
\end{deluxetable*}

\subsection{Error Analysis}
\label{erroranalysis}

I determined the values of $\Gamma$ and $\theta$ and their errors from
a Monte Carlo (MC) error analysis.  For each MC iteration, I randomly
drew:
\begin{itemize}
\item a core shift, $\Delta\phi$, based on the measured value and
  assuming a normally distributed error of 51 $\mu$asec.
\item an apparent jet opening angle, $\alpha_{\rm app}$, based on the
  measured value and assuming a normally distributed error of 10\% of
  the measured value.
\item values of $c_1$ and $c_2$ assuming values and normally
  distributed errors $c_1=0.64\pm0.09$ and $c_2=43.54\pm0.12$
  \citep{cavagnolo10}.  From the randomly drawn $c_1$, $c_2$, and
  $L_{\rm ext}$, the jet power $P_j$ is calculated from Equation
  (\ref{powercavag}).
\item values of the nuisance parameters were all drawn from flat
  priors with limits as follows:
\begin{itemize}
\item $p$ between 1 and 5.
\item $\log_{10}\g_1$ between 0 and 4.
\item $\log_{10}\g_2$ between 3 and 7.
\item $\log_{10}\xi_e$ between -2 and  2.
\item $\log_{10}\xi_p$ between -4 and 2.
\item $\log_{10}\xi_m$ between -4 and 2.
\end{itemize}
\end{itemize}
If the randomly drawn $\g_2$ was less than the randomly drawn $\g_1$,
I redrew both parameters.  The th parameters $\xi_p$ and
  $\xi_m$ were allowed to go much lower than $\xi_e$ to
include the possibility of an electron-positron jet with few protons,
accelerated or otherwise.  Note that this
  marginalizes over the widely-used magnetization parameter,
\begin{flalign}
\label{magnetization}
\sigma & \equiv \frac{g_B \hat\g_{\rm ad} u_B}{u_e + u_p + \rho c^2} 
\nonumber \\ 
& = \frac{g_B\hat\g_{\rm ad}}{\xi_e + \xi_p + \hat\g_{\rm ad}\xi_m}\ ,
\end{flalign}
between $\sigma=0.006$ and $\sigma=200$.

Once all of the parameters are randomly drawn, the values $\G$ and
$\theta$ are computed by solving Equations (\ref{dphi2}) and
(\ref{fsy3}) numerically.  Then other jet parameters, $\alpha$, $B(1
\pc)$, $\dD$, and $\beta_{\rm app}$ are computed from these values.
This is repeated for $N=10^4$ iterations.  I compute the magnetic
field at $r=1\ \pc$ to be consistent with other calculations of this
quantity \citep[e.g.,][]{pushkarev12,zdz15}.  The statistical
properties of the results can then be computed from these iterations.

\subsection{Jet Parameters}

\begin{figure*}
\epsscale{1.0} 
\plotone{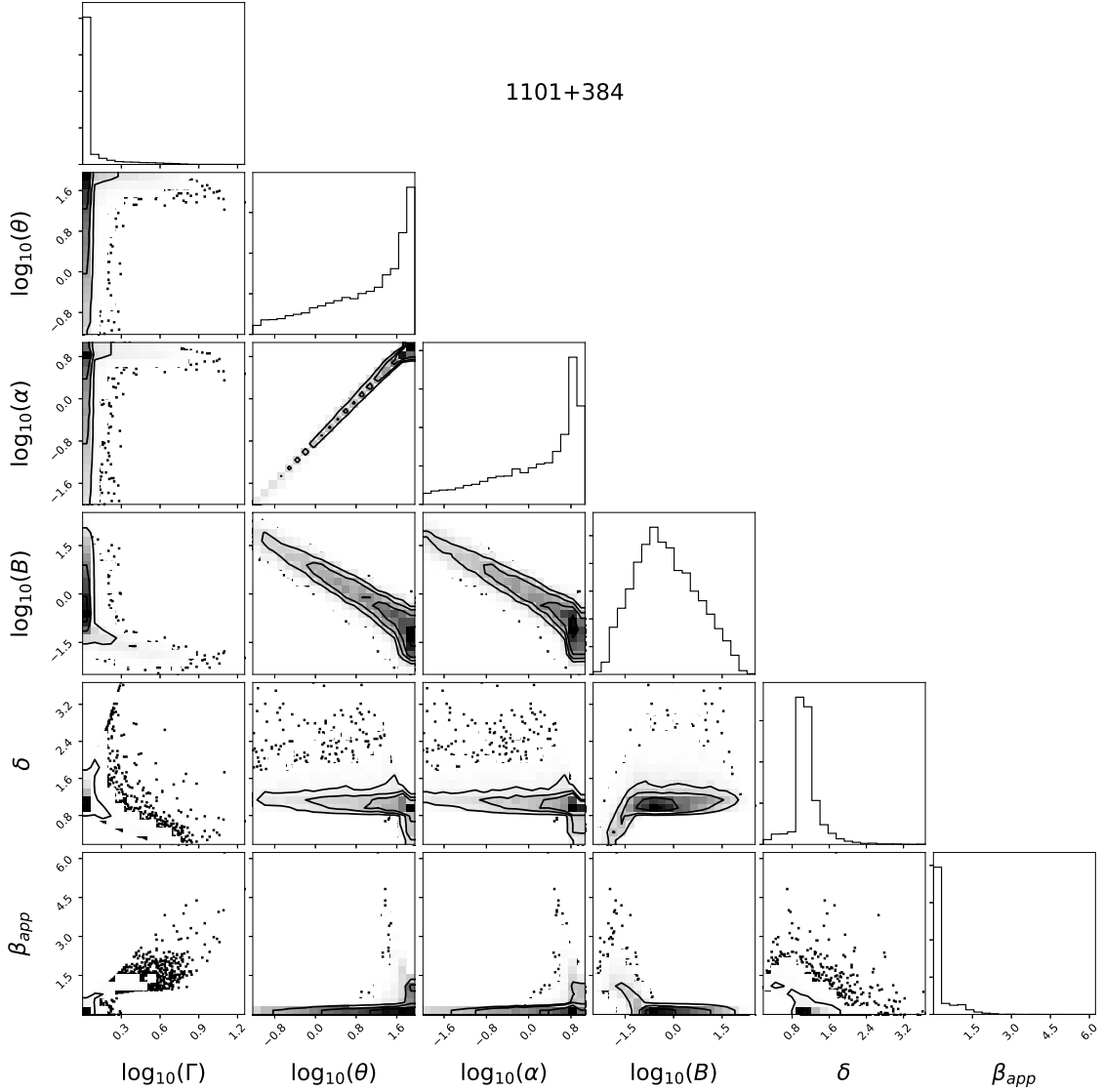}
\caption{Result of MC calculation for 1101+384 (Mrk 421).  This plot
  was made with {\em corner.py} \citep{foreman16}.}
\label{1101MC}
\end{figure*}

\begin{figure*}
\epsscale{1.0} 
\plotone{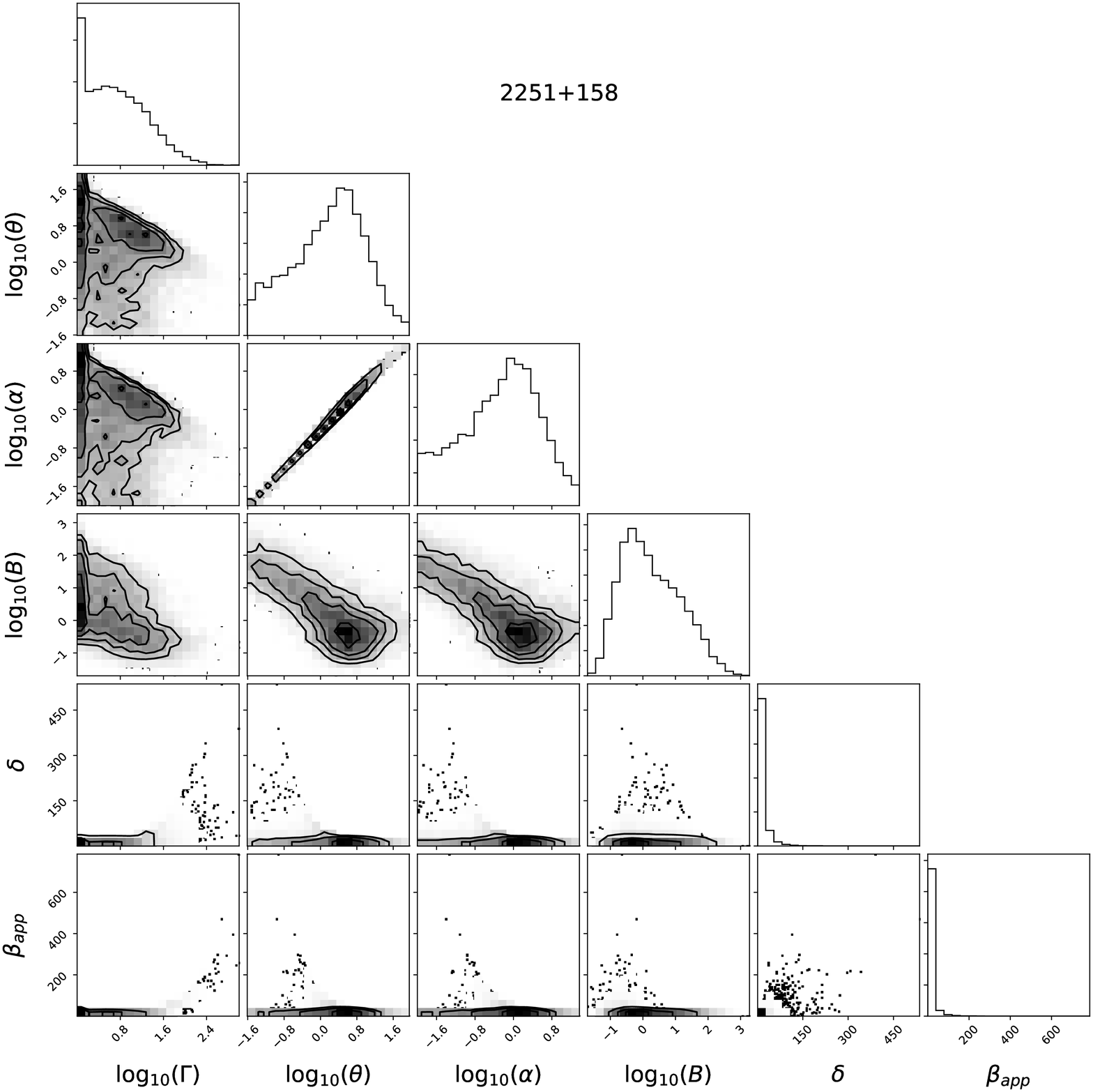}
\caption{Result of MC calculation for 2251+158 (3C 454.3).  This plot
  was made with {\em corner.py} \citep{foreman16}.}
\label{2251MC}
\vspace{5.2mm}
\end{figure*}

The results for two MC calculations can be seen in Figures
\ref{1101MC} and \ref{2251MC}.  Figure \ref{1101MC} shows my result
for the famous nearby BL Lac object 1101+384 (Mrk 421).  The
parameters $\G$ and $\dD$ are well-constrained to low values, while
$\theta$ is poorly constrained.  It is interesting to compare the
small values of $\G$ and $\dD$ here with the values inferred from the
superluminal components seen in VLBI images
\citep{piner04,piner05,piner08,piner10}.  This is discussed further in
Section \ref{gammaraysection}.  Many of the parameters are
well-correlated, such as $\theta$ and $\alpha$, $\theta$ and
$B(1\ \pc)$, and $B(1\ \pc)$ and $\alpha$.

Figure \ref{2251MC} shows my result for the famous $\g$-ray bright
FSRQ 2251+158 (3C 454.3).  The angle $\theta$ is much more strongly
constrained than for 1101+384, while $\G$ and $\dD$ are much more
poorly constrained.  The same parameter correlations seen for 1101+384
are seen for 2251+158 as well, but it is also clear that $\G$ and
$\theta$ are well correlated also.

I present the median and 68\% confidence intervals of my MC
calculation in Table \ref{tableresult}.  Errors are quite large in
many cases, but in some cases parameters are well constrained.  For
instance, as mentioned above, $\G$ and $\dD$ are well-constrained for
1101+384.  Generally $\dD$ is well-constrained for low $\dD$ sources,
while $\theta$ is more strongly constrained for higher $\dD$ sources.

\subsection{Alternative Jet Power Calculation}
\label{altjetpower}

To estimate the jet power of the sources in my sample (Section
\ref{jetpowersection}), I used the relation between radio lobe
luminosity and jet power found by \citet[][]{cavagnolo10}.  This relation
was found to be in agreement with the theoretical prediction of
\citet{willott99} made for Fanaroff-Riley (FR) type II radio galaxies.
They used narrow-line luminosities as a proxy for jet power, and found
an empirical relation that agreed with their model prediction.
Similar empirical correlations have been found for FR I
\citep[e.g.,][]{birzan04,birzan08,cavagnolo10,osullivan11} and FR II
\citep[e.g.,][]{daly12} radio galaxies, with authors measuring the jet
power in a variety of ways.  Agreement between this correlation for FR
II and FR I sources was found, contrary to theoretical expectations
\citep{godfrey13}.  \citet{godfrey16} pointed out that these empirical
correlations were probably the result of both the radio luminosity and
jet power being dependent on source distance.  When taking into
account this effect, \citet{ineson17} found a correlation
\begin{flalign}
\label{powerineson}
\log_{10}\left[\frac{P_j}{\erg\ \s^{-1}}\right] = 
c_3 \left\{\log_{10}\left[\frac{L_{\rm ext}}{\erg\ \s^{-1}}\right]-44.3\right\} +
c_4\ ,
\end{flalign}
with $c_3=0.89\pm0.09$, and $c_4=46.7$, in agreement with
the theoretical prediction by \citet{willott99} for FR II sources.

\begin{figure*}
\vspace{10.0mm} 
\epsscale{1.0} 
\plottwo{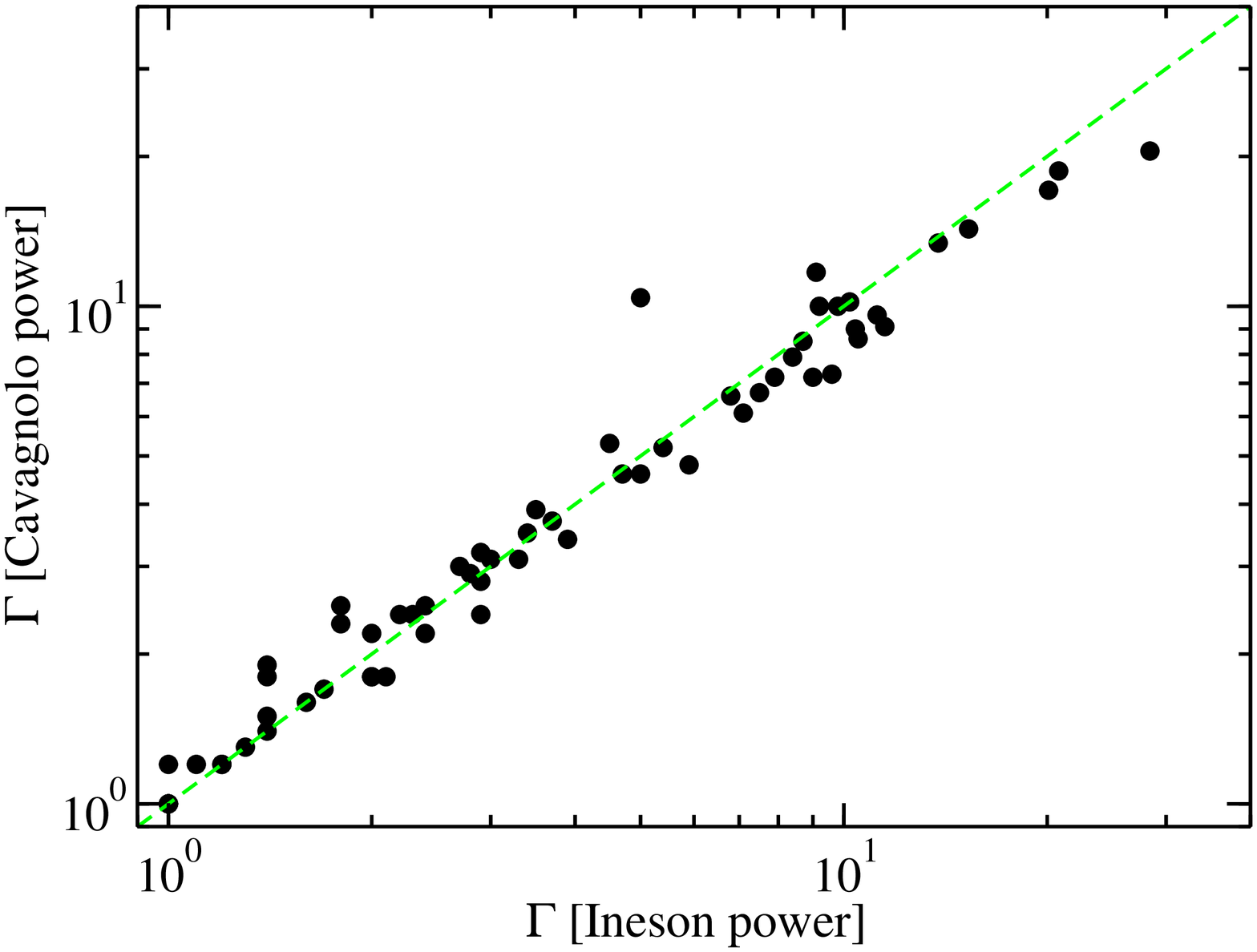}{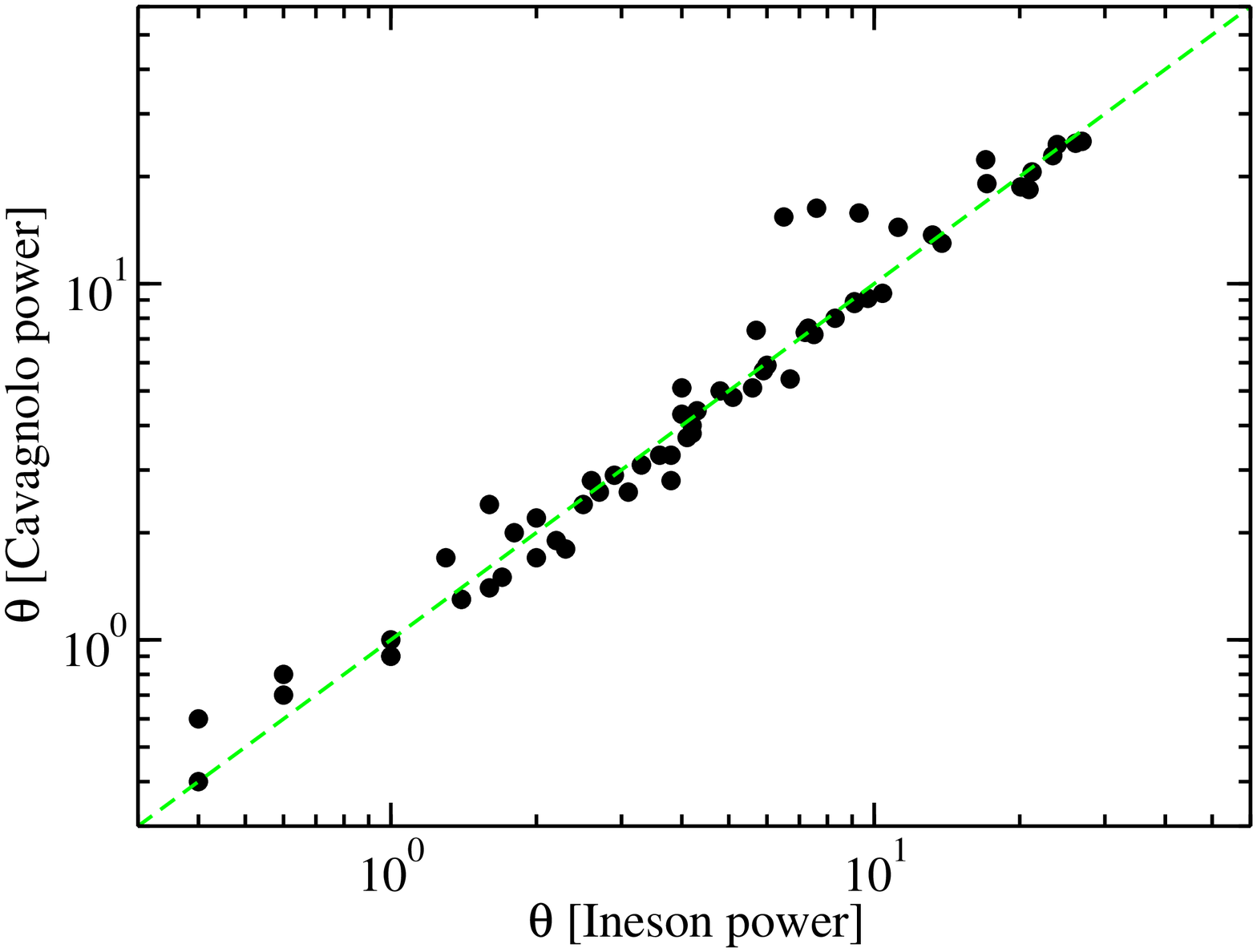}
\caption{A comparison of results computed using the relation between
  $L_{\rm ext}$ and $P_J$ found by \citet[][Equation
    (\ref{powercavag})]{cavagnolo10} and \citet[][Equation
    (\ref{powerineson})]{ineson17}.  Left: A comparison of the bulk
  Lorentz factor $\G$.  Right: A comparison of the angle to the line
  of sight $\theta$.  Error bars are not plotted for clarity.  The
  dashed lines indicates the where the results with the two different
  calculations are equal.}
\label{powercompare}
\vspace{2.2mm}
\end{figure*}

In Figure \ref{powercompare} I compare my results computed with the
relation found by \citet[][Equation (\ref{powercavag})]{cavagnolo10}
and \citet[][Equation (\ref{powerineson})]{ineson17} where in
  both cases I divide the power by 2 to account for only a single
  jet.  In all cases, the results are within the errors of each other
(error bars are not shown on the plot).  I conclude that the jet power
relation used has little effect on my results.

\LongTables
\begin{deluxetable*}{llcccccc}
\tabletypesize{\scriptsize}
\tablecaption{Blazar Jet Parameter Results}
\tablewidth{0pt}
\tablehead{
\colhead{Source} &
\colhead{Alias} &
\colhead{$\G$} &
\colhead{$\theta$ [\arcdeg]} &
\colhead{$\alpha$ [\arcdeg]} &
\colhead{$B(1\ \pc)$ [G]} &
\colhead{$\delta_D$} &
\colhead{$\beta_{\rm app}$ } 
}
\startdata
0133$+$476 &  DA 55         &   $   3.0^{+  13.5}_{- 1.9} $&$  1.7^{+  9.7}_{-  1.5} $&$  0.3^{+ 1.9}_{- 0.3} $&$  1.2^{+ 12.1}_{- 1.1} $&$  4.8^{+  17.0}_{-   3.5} $&$  0.4^{+   9.4}_{-  0.4}$  \\ 
0202$+$149 &  4C +15.05     &	$   2.3^{+   7.1}_{- 1.2} $&$  5.1^{+ 20.2}_{-  4.6} $&$  0.7^{+ 2.8}_{- 0.6} $&$  0.5^{+  5.6}_{- 0.4} $&$  3.2^{+   7.5}_{-   1.9} $&$  0.7^{+   5.9}_{-  0.6}$  \\ 
0212$+$735 &  S5 0212$+$73  &	$   3.7^{+  15.2}_{- 2.7} $&$  0.8^{+  4.2}_{-  0.7} $&$  0.1^{+ 0.6}_{- 0.1} $&$  4.7^{+ 26.5}_{- 4.3} $&$  6.8^{+  20.5}_{-   5.5} $&$  0.3^{+  11.0}_{-  0.3}$  \\ 
0215$+$015 &  OD 026        &	$   2.4^{+   7.7}_{- 1.3} $&$  5.4^{+ 21.8}_{-  5.0} $&$  1.7^{+ 6.7}_{- 1.6} $&$  0.9^{+ 13.4}_{- 0.7} $&$  3.2^{+   7.7}_{-   1.9} $&$  0.8^{+   6.1}_{-  0.8}$  \\ 
0234$+$285 &  4C 28.07      &	$   1.8^{+   3.2}_{- 0.7} $&$  9.4^{+ 22.8}_{-  8.7} $&$  1.6^{+ 3.7}_{- 1.5} $&$  1.5^{+ 19.1}_{- 1.2} $&$  2.3^{+   3.1}_{-   1.1} $&$  0.5^{+   3.2}_{-  0.5}$  \\ 
0333$+$321 &  NRAO 140      &	$   2.2^{+   4.3}_{- 1.1} $&$ 13.0^{+ 24.3}_{- 11.0} $&$  0.9^{+ 1.5}_{- 0.8} $&$  1.5^{+ 12.0}_{- 1.1} $&$  2.4^{+   3.2}_{-   1.2} $&$  1.1^{+   3.9}_{-  1.0}$  \\ 
0336$-$019 &  4C 28.07      &	$   3.1^{+  12.8}_{- 2.0} $&$  2.0^{+ 10.7}_{-  1.8} $&$  0.5^{+ 2.5}_{- 0.4} $&$  1.1^{+ 12.0}_{- 0.9} $&$  4.9^{+  15.3}_{-   3.5} $&$  0.6^{+   9.2}_{-  0.5}$  \\ 
0403$-$132 &  PKS 0403$-$13 &	$   1.8^{+   2.4}_{- 0.7} $&$ 24.8^{+ 32.1}_{- 22.4} $&$  3.4^{+ 3.5}_{- 3.0} $&$  0.8^{+  8.2}_{- 0.5} $&$  1.5^{+   1.6}_{-   0.6} $&$  1.0^{+   2.0}_{-  0.9}$  \\ 
0420$-$014 &  PKS 0420$-$01 &	$   1.4^{+   1.8}_{- 0.3} $&$  8.0^{+ 25.8}_{-  7.5} $&$  1.6^{+ 4.8}_{- 1.5} $&$  1.9^{+ 25.1}_{- 1.7} $&$  1.8^{+   2.1}_{-   0.7} $&$  0.1^{+   1.9}_{-  0.1}$  \\ 
0528$+$134 &  PKS 0528$+$134&	$   9.1^{+  35.7}_{- 7.2} $&$  1.7^{+  5.4}_{-  1.4} $&$  0.2^{+ 0.8}_{- 0.2} $&$  1.8^{+ 13.4}_{- 1.3} $&$ 12.2^{+  29.2}_{-   9.1} $&$  4.3^{+  30.6}_{-  4.1}$  \\ 
0605$-$085 &  OC $-$010     &	$   7.9^{+  41.1}_{- 6.2} $&$  3.1^{+ 13.0}_{-  2.6} $&$  0.4^{+ 1.6}_{- 0.3} $&$  0.6^{+  4.0}_{- 0.4} $&$  8.2^{+  28.9}_{-   6.1} $&$  4.9^{+  31.8}_{-  4.5}$  \\ 
0607$-$157 &  PKS 0607$-$15 &	$   1.0^{+   0.0}_{- 0.0} $&$  2.8^{+ 26.3}_{-  2.7} $&$  0.9^{+ 7.6}_{- 0.8} $&$  3.1^{+ 33.6}_{- 2.8} $&$  1.0^{+   0.2}_{-   0.0} $&$  0.0^{+   0.0}_{-  0.0}$  \\ 
0716$+$714 &  S5 0716$+$71  &	$   2.4^{+   6.0}_{- 1.3} $&$ 13.7^{+ 30.4}_{- 11.8} $&$  2.0^{+ 3.9}_{- 1.7} $&$  0.3^{+  2.3}_{- 0.2} $&$  2.3^{+   4.1}_{-   1.2} $&$  1.4^{+   4.7}_{-  1.3}$  \\ 
0738$+$313 &  OI 363        &	$   3.9^{+  13.2}_{- 2.6} $&$  5.0^{+ 15.3}_{-  4.2} $&$  0.5^{+ 1.4}_{- 0.4} $&$  0.7^{+  5.8}_{- 0.5} $&$  4.8^{+  10.8}_{-   3.2} $&$  1.9^{+  11.2}_{-  1.8}$  \\ 
0748$+$126 &  OI 280	  &     $  13.4^{+  74.6}_{-11.2} $&$  1.0^{+  4.3}_{-  0.8} $&$  0.1^{+ 0.6}_{- 0.1} $&$  0.9^{+  6.2}_{- 0.7} $&$ 18.1^{+  66.2}_{-  14.6} $&$  6.3^{+  64.3}_{-  6.1}$  \\ 
0754$+$100 &  PKS 0754$+$100&	$   1.2^{+   1.0}_{- 0.2} $&$ 22.3^{+ 37.5}_{- 21.2} $&$  2.5^{+ 3.3}_{- 2.4} $&$  0.6^{+ 11.2}_{- 0.5} $&$  1.3^{+   0.9}_{-   0.3} $&$  0.3^{+   1.2}_{-  0.3}$  \\ 
0804$+$499 &  OJ 508        &	$   2.2^{+   9.7}_{- 1.1} $&$  2.8^{+ 17.2}_{-  2.5} $&$  0.8^{+ 5.2}_{- 0.8} $&$  0.9^{+ 12.0}_{- 0.8} $&$  3.2^{+  11.9}_{-   2.0} $&$  0.3^{+   6.4}_{-  0.3}$  \\ 
0805$-$077 &  PKS 0805$-$07 &	$   1.8^{+   2.7}_{- 0.7} $&$ 18.4^{+ 30.8}_{- 16.8} $&$  2.9^{+ 4.2}_{- 2.6} $&$  1.3^{+ 15.9}_{- 1.0} $&$  1.8^{+   2.1}_{-   0.7} $&$  0.8^{+   2.4}_{-  0.8}$  \\ 
0823$+$033 &  PKS 0823$+$033&	$   1.9^{+   5.6}_{- 0.9} $&$  2.4^{+ 11.6}_{-  2.1} $&$  0.3^{+ 1.4}_{- 0.2} $&$  1.4^{+ 11.1}_{- 1.3} $&$  3.1^{+   7.3}_{-   2.0} $&$  0.2^{+   4.2}_{-  0.2}$  \\ 
0827$+$243 &  OJ 248        &	$   2.5^{+   6.4}_{- 1.3} $&$  8.8^{+ 23.6}_{-  7.8} $&$  1.1^{+ 2.9}_{- 1.0} $&$  0.7^{+  7.8}_{- 0.5} $&$  2.9^{+   5.4}_{-   1.6} $&$  1.1^{+   5.4}_{-  1.1}$  \\ 
0829$+$046 &  OJ 049        &	$   1.8^{+   3.3}_{- 0.7} $&$ 14.4^{+ 31.6}_{- 13.1} $&$  2.3^{+ 4.4}_{- 2.1} $&$  0.2^{+  2.7}_{- 0.1} $&$  1.9^{+   2.8}_{-   0.8} $&$  0.7^{+   2.8}_{-  0.7}$  \\ 
0836$+$710 &  4C +71.07     &	$   7.3^{+  25.1}_{- 5.5} $&$  3.3^{+  9.4}_{-  2.6} $&$  0.4^{+ 1.0}_{- 0.3} $&$  1.7^{+ 11.3}_{- 1.2} $&$  8.2^{+  16.9}_{-   5.7} $&$  4.3^{+  20.8}_{-  4.0}$  \\ 
0906$+$015 &  4C +01.24     &   $   1.7^{+   3.0}_{- 0.6} $&$  8.9^{+ 24.2}_{-  8.2} $&$  1.3^{+ 3.5}_{- 1.2} $&$  1.3^{+ 16.3}_{- 1.1} $&$  2.2^{+   3.1}_{-   1.0} $&$  0.4^{+   2.9}_{-  0.4}$  \\ 
0917$+$624 &  OK 630        &   $   4.6^{+  18.7}_{- 3.3} $&$  3.8^{+ 14.6}_{-  3.3} $&$  0.5^{+ 2.0}_{- 0.4} $&$  0.9^{+  8.5}_{- 0.7} $&$  5.6^{+  15.9}_{-   3.9} $&$  2.2^{+  15.2}_{-  2.1}$  \\ 
0945$+$408 &  4C +40.24     &	$   7.2^{+  31.7}_{- 5.4} $&$  4.8^{+ 16.7}_{-  3.9} $&$  0.6^{+ 2.0}_{- 0.5} $&$  0.8^{+  4.0}_{- 0.6} $&$  6.4^{+  18.2}_{-   4.5} $&$  4.9^{+  23.5}_{-  4.4}$  \\ 
1038$+$064 &  4C +06.41     &	$   8.5^{+  32.6}_{- 6.6} $&$  3.1^{+  9.3}_{-  2.4} $&$  0.2^{+ 0.5}_{- 0.1} $&$  1.2^{+  6.2}_{- 0.9} $&$  9.1^{+  21.5}_{-   6.6} $&$  5.6^{+  26.4}_{-  5.2}$  \\ 
1045$-$188 &  PKS 1045$-$18 &	$   7.2^{+  23.7}_{- 5.2} $&$  7.2^{+ 18.0}_{-  5.4} $&$  0.5^{+ 1.2}_{- 0.4} $&$  0.8^{+  2.5}_{- 0.5} $&$  4.9^{+  10.5}_{-   3.2} $&$  5.2^{+  16.0}_{-  4.3}$  \\ 
1101$+$384 &  Mrk 421       &	$   1.0^{+   0.2}_{- 0.0} $&$ 16.3^{+ 49.3}_{- 15.4} $&$  2.5^{+ 5.3}_{- 2.4} $&$  0.5^{+  6.0}_{- 0.4} $&$  1.1^{+   0.2}_{-   0.1} $&$  0.0^{+   0.5}_{-  0.0}$  \\ 
1127$-$145 &  PKS 1127$-$14 &	$  18.7^{+ 112.0}_{-16.0} $&$  0.9^{+  4.4}_{-  0.7} $&$  0.1^{+ 0.6}_{- 0.1} $&$  0.9^{+  5.6}_{- 0.7} $&$ 22.6^{+  88.0}_{-  18.5} $&$ 10.8^{+  93.2}_{- 10.4}$  \\ 
1150$+$812 &  S5 1150$+$81  &	$  17.1^{+ 106.5}_{-14.6} $&$  1.4^{+  7.4}_{-  1.2} $&$  0.2^{+ 1.0}_{- 0.2} $&$  0.8^{+  4.2}_{- 0.6} $&$ 17.3^{+  72.5}_{-  14.0} $&$ 10.9^{+  86.6}_{- 10.3}$  \\ 
1156$+$295 &  4C +29.45     &	$   5.2^{+  17.6}_{- 3.8} $&$  3.3^{+ 10.3}_{-  2.8} $&$  0.5^{+ 1.5}_{- 0.4} $&$  0.9^{+  8.4}_{- 0.7} $&$  6.7^{+  14.7}_{-   4.6} $&$  2.3^{+  15.0}_{-  2.2}$  \\ 
1219$+$044 &  4C +04.42     &	$   2.8^{+   7.2}_{- 1.6} $&$  9.1^{+ 21.7}_{-  7.8} $&$  1.0^{+ 2.3}_{- 0.9} $&$  0.9^{+  7.7}_{- 0.6} $&$  3.1^{+   5.5}_{-   1.8} $&$  1.5^{+   6.1}_{-  1.3}$  \\ 
1222$+$216 &  4C +21.35     &	$   3.1^{+   6.7}_{- 1.7} $&$ 20.6^{+ 31.0}_{- 15.5} $&$  1.9^{+ 2.4}_{- 1.4} $&$  0.5^{+  1.5}_{- 0.3} $&$  1.9^{+   2.9}_{-   0.9} $&$  2.0^{+   4.2}_{-  1.5}$  \\ 
1308$+$326 &  OP 313        &	$   4.6^{+  21.6}_{- 3.3} $&$  4.0^{+ 17.0}_{-  3.5} $&$  0.6^{+ 2.7}_{- 0.6} $&$  0.6^{+  5.7}_{- 0.5} $&$  5.5^{+  17.9}_{-   3.8} $&$  2.3^{+  16.5}_{-  2.2}$  \\ 
1334$-$127 &  PKS 1335$-$127&	$   2.4^{+   5.2}_{- 1.3} $&$  7.3^{+ 16.8}_{-  6.4} $&$  0.8^{+ 1.8}_{- 0.7} $&$  1.1^{+ 11.7}_{- 0.8} $&$  3.2^{+   4.6}_{-   1.8} $&$  0.8^{+   5.2}_{-  0.8}$  \\ 
1413$+$135 &  PKS B1413$+$135&	$   1.2^{+   1.1}_{- 0.2} $&$ 15.8^{+ 34.8}_{- 14.6} $&$  1.2^{+ 2.2}_{- 1.1} $&$  0.6^{+  6.9}_{- 0.5} $&$  1.4^{+   1.2}_{-   0.3} $&$  0.2^{+   1.3}_{-  0.2}$  \\ 
1502$+$106 &  OR 103	   &    $   8.6^{+  91.1}_{- 7.3} $&$  0.9^{+  7.6}_{-  0.7} $&$  0.3^{+ 2.5}_{- 0.2} $&$  0.8^{+  8.8}_{- 0.6} $&$ 12.5^{+  96.6}_{-  10.6} $&$  3.0^{+  66.7}_{-  3.0}$  \\ 
1504$-$166 &  PKS 1504$-$167&	$   2.9^{+  10.0}_{- 1.8} $&$  4.4^{+ 17.2}_{-  3.9} $&$  0.7^{+ 2.7}_{- 0.6} $&$  0.8^{+  8.7}_{- 0.6} $&$  4.0^{+  10.2}_{-   2.6} $&$  0.9^{+   8.2}_{-  0.9}$  \\ 
1510$-$089 &  PKS 1510$-$08 &	$   3.2^{+   9.1}_{- 1.9} $&$  7.5^{+ 20.1}_{-  6.4} $&$  1.0^{+ 2.6}_{- 0.8} $&$  0.4^{+  3.9}_{- 0.3} $&$  3.6^{+   7.1}_{-   2.2} $&$  1.6^{+   7.6}_{-  1.5}$  \\ 
1538$+$149 &  4C +14.60     &	$  10.0^{+  65.7}_{- 8.1} $&$  2.6^{+ 13.2}_{-  2.2} $&$  0.4^{+ 1.8}_{- 0.3} $&$  0.4^{+  2.4}_{- 0.3} $&$  9.8^{+  44.2}_{-   7.7} $&$  6.3^{+  50.1}_{-  5.8}$  \\ 
1606$+$106 &  4C +10.45     &	$   6.6^{+  47.1}_{- 5.3} $&$  1.3^{+  8.7}_{-  1.1} $&$  0.3^{+ 1.8}_{- 0.2} $&$  0.8^{+  8.0}_{- 0.6} $&$  9.3^{+  49.7}_{-   7.4} $&$  2.3^{+  35.8}_{-  2.3}$  \\ 
1611$+$343 &  DA 406        &	$  14.3^{+ 149.4}_{-12.8} $&$  0.4^{+  2.7}_{-  0.3} $&$  0.1^{+ 0.6}_{- 0.1} $&$  1.1^{+ 10.8}_{- 0.9} $&$ 23.1^{+ 167.2}_{-  20.6} $&$  3.8^{+ 121.4}_{-  3.8}$  \\ 
1633$+$382 &  4C +38.41     &	$   3.4^{+  11.6}_{- 2.3} $&$  2.6^{+ 10.6}_{-  2.3} $&$  0.5^{+ 2.1}_{- 0.5} $&$  1.7^{+ 19.2}_{- 1.4} $&$  5.2^{+  12.6}_{-   3.6} $&$  0.8^{+   9.7}_{-  0.8}$  \\ 
1637$+$574 &  OS 562        &	$  10.2^{+  50.5}_{- 8.2} $&$  2.4^{+  9.4}_{-  1.9} $&$  0.2^{+ 0.9}_{- 0.2} $&$  0.7^{+  3.8}_{- 0.5} $&$ 10.8^{+  35.1}_{-   8.2} $&$  6.5^{+  40.5}_{-  6.0}$  \\ 
1641$+$399 &  3C 345        &	$   6.7^{+  21.4}_{- 5.0} $&$  3.7^{+  9.7}_{-  3.0} $&$  0.4^{+ 1.1}_{- 0.3} $&$  0.9^{+  6.3}_{- 0.7} $&$  7.5^{+  13.8}_{-   5.1} $&$  4.0^{+  17.8}_{-  3.7}$  \\ 
1652$+$398 &  Mrk 501       &	$   1.0^{+   0.4}_{- 0.0} $&$ 15.4^{+ 47.8}_{- 14.8} $&$  2.5^{+ 5.8}_{- 2.4} $&$  0.3^{+  6.4}_{- 0.3} $&$  1.1^{+   0.4}_{-   0.1} $&$  0.0^{+   0.8}_{-  0.0}$  \\ 
1730$-$130 &  NRAO 530      &	$   6.1^{+  19.0}_{- 4.4} $&$  5.1^{+ 13.0}_{-  4.0} $&$  0.5^{+ 1.2}_{- 0.4} $&$  1.1^{+  6.3}_{- 0.8} $&$  6.0^{+  11.4}_{-   3.9} $&$  4.0^{+  14.7}_{-  3.6}$  \\ 
1749$+$096 &  4C +09.57     &	$  11.7^{+  78.2}_{- 9.9} $&$  0.7^{+  3.4}_{-  0.5} $&$  0.1^{+ 0.5}_{- 0.1} $&$  0.5^{+  4.0}_{- 0.4} $&$ 17.9^{+  80.8}_{-  15.0} $&$  4.0^{+  66.9}_{-  4.0}$  \\ 
1807$+$698 &  3C 371        &	$   2.5^{+   4.8}_{- 1.3} $&$ 19.1^{+ 29.1}_{- 15.1} $&$  1.8^{+ 2.3}_{- 1.4} $&$  0.1^{+  0.6}_{- 0.1} $&$  2.0^{+   2.8}_{-   0.9} $&$  1.6^{+   3.6}_{-  1.3}$  \\ 
1928$+$738 &  4C +73.18     &	$  10.0^{+  38.1}_{- 7.8} $&$  2.9^{+  8.5}_{-  2.2} $&$  0.2^{+ 0.7}_{- 0.2} $&$  0.5^{+  2.6}_{- 0.3} $&$ 10.2^{+  23.0}_{-   7.3} $&$  6.8^{+  30.4}_{-  6.3}$  \\ 
1936$-$155 &  PKS 1936$-$15 &	$   1.0^{+   0.1}_{- 0.0} $&$  7.4^{+ 42.6}_{-  7.1} $&$  2.3^{+10.8}_{- 2.2} $&$  7.0^{+111.0}_{- 6.4} $&$  1.0^{+   0.3}_{-   0.0} $&$  0.0^{+   0.2}_{-  0.0}$  \\ 
2121$+$053 &  PKS 2121$+$053&	$   1.3^{+   2.4}_{- 0.3} $&$  2.2^{+ 16.7}_{-  2.1} $&$  0.7^{+ 4.8}_{- 0.6} $&$  3.9^{+ 38.9}_{- 3.6} $&$  1.9^{+   3.9}_{-   0.9} $&$  0.0^{+   1.3}_{-  0.0}$  \\ 
2128$-$123 &  PKS 2128$-$12 &	$   5.3^{+  15.6}_{- 3.8} $&$  4.3^{+  9.1}_{-  3.2} $&$  0.2^{+ 0.4}_{- 0.1} $&$  1.1^{+  6.3}_{- 0.8} $&$  6.5^{+  10.5}_{-   4.3} $&$  3.0^{+  14.1}_{-  2.8}$  \\ 
2131$-$021 &  4C $−$02.81   &	$   9.0^{+  46.6}_{- 7.3} $&$  1.9^{+  8.6}_{-  1.6} $&$  0.3^{+ 1.4}_{- 0.3} $&$  0.9^{+  7.2}_{- 0.6} $&$ 10.7^{+  38.0}_{-   8.2} $&$  4.7^{+  37.4}_{-  4.5}$  \\ 
2134$+$004 &  PKS 2134+004  &	$   9.6^{+  34.0}_{- 7.6} $&$  1.3^{+  3.6}_{-  1.0} $&$  0.2^{+ 0.5}_{- 0.1} $&$  2.3^{+ 17.1}_{- 1.8} $&$ 14.0^{+  30.6}_{-  10.5} $&$  3.6^{+  31.8}_{-  3.5}$  \\ 
2155$-$152 &  PKS 2155$-$152&	$   1.2^{+   1.0}_{- 0.2} $&$ 22.9^{+ 37.3}_{- 22.0} $&$  3.3^{+ 4.2}_{- 3.2} $&$  1.3^{+ 29.5}_{- 1.1} $&$  1.3^{+   0.9}_{-   0.3} $&$  0.3^{+   1.2}_{-  0.3}$  \\ 
2200$+$420 &  BL Lac        &	$  10.4^{+ 118.8}_{- 9.0} $&$  0.6^{+  4.8}_{-  0.5} $&$  0.1^{+ 1.1}_{- 0.1} $&$  0.1^{+  1.2}_{- 0.1} $&$ 15.6^{+ 128.6}_{-  13.5} $&$  3.2^{+  90.7}_{-  3.1}$  \\ 
2201$+$171 &  PKS 2201$+$171&	$   1.2^{+   1.0}_{- 0.2} $&$ 25.1^{+ 37.9}_{- 24.1} $&$  2.8^{+ 3.1}_{- 2.7} $&$  2.1^{+ 44.5}_{- 1.7} $&$  1.2^{+   0.7}_{-   0.2} $&$  0.2^{+   1.1}_{-  0.2}$  \\ 
2201$+$315 &  4C +31.63     &	$   1.5^{+   1.6}_{- 0.5} $&$ 24.6^{+ 32.0}_{- 22.9} $&$  2.6^{+ 2.7}_{- 2.4} $&$  0.6^{+  8.4}_{- 0.4} $&$  1.5^{+   1.3}_{-   0.4} $&$  0.6^{+   1.6}_{-  0.6}$  \\ 
2223$-$052 &  3C 446        &	$  20.5^{+  80.9}_{-16.7} $&$  1.5^{+  4.5}_{-  1.1} $&$  0.2^{+ 0.5}_{- 0.1} $&$  1.5^{+  6.4}_{- 1.0} $&$ 19.6^{+  43.1}_{-  14.4} $&$ 14.9^{+  62.9}_{- 13.7}$  \\ 
2227$-$088 &  PHL 5225      &	$   1.6^{+   3.1}_{- 0.6} $&$  5.9^{+ 19.9}_{-  5.4} $&$  0.8^{+ 2.7}_{- 0.7} $&$  2.2^{+ 23.7}_{- 1.9} $&$  2.4^{+   3.7}_{-   1.2} $&$  0.2^{+   2.9}_{-  0.2}$  \\ 
2230$+$114 &  CTA 102       &	$   1.8^{+   2.7}_{- 0.7} $&$ 18.7^{+ 28.9}_{- 16.7} $&$  2.1^{+ 2.8}_{- 1.9} $&$  1.3^{+ 13.1}_{- 1.0} $&$  1.8^{+   2.0}_{-   0.7} $&$  0.9^{+   2.5}_{-  0.8}$  \\ 
2251$+$158 &  3C 454.3      &	$   4.8^{+  16.1}_{- 3.5} $&$  1.8^{+  7.1}_{-  1.7} $&$  0.7^{+ 2.6}_{- 0.6} $&$  1.3^{+ 16.2}_{- 1.1} $&$  7.3^{+  16.9}_{-   5.2} $&$  1.1^{+  13.7}_{-  1.0}$  \\ 
2345$-$167 &  PKS 2345$-$16 &	$   3.5^{+  10.4}_{- 2.2} $&$  5.7^{+ 16.2}_{-  4.9} $&$  0.8^{+ 2.2}_{- 0.7} $&$  0.7^{+  6.5}_{- 0.5} $&$  4.2^{+   8.7}_{-   2.7} $&$  1.6^{+   9.0}_{-  1.5}$     
\enddata
\label{tableresult}
\end{deluxetable*}

\section{Implications}
\label{implications}

I explore some implications of my results in this section.  I start
out by comparing my results with previous calculations of jet
parameters (Section \ref{compareprevious}).  I then explore what
observables could be a good proxy for $\theta$ (Section
\ref{jetanglesection}), what parsec-scale jet parameters could be
related to $\g$-ray emission (Section \ref{gammaraysection}), and
implications for jet physics (Section \ref{jetphysics}).  To do this,
I test correlations between different parameters and results three
different ways.  I used the non-parametric Spearman and Kendall rank
correlation tests to determine the significance of a correlation
between parameters.  I also perform linear fits of the form:
\begin{eqnarray}
y = m(x-x_n) + y_n\ .
\end{eqnarray}
I performed two fits for each set of variables: with $m$ and $y_n$ both
as free parameters, and with only $b$ as a free parameter, fixing
$m=0$.  I then used an F-test to determine the significance of the
model with $m$ as a free parameter versus the model with it fixed.
The significance of these tests and resulting $m$ and $y_n$ from the
fits with both as free parameters can be found in Table
\ref{table_correlate}.

\subsection{Comparison with Previous Results}
\label{compareprevious}

\begin{figure}
\vspace{10.0mm} 
\epsscale{1.0} 
\plotone{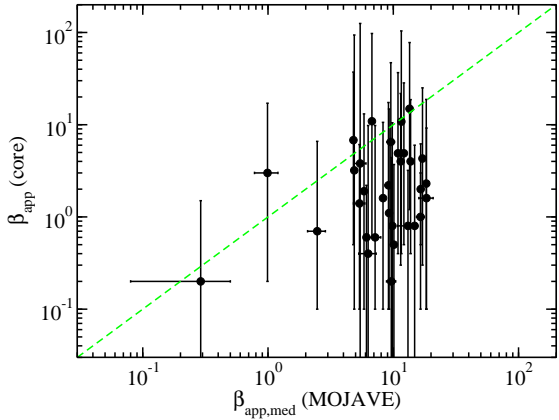}
\caption{The apparent jet speed of the core from this work plotted
  against the median apparent speed of jet components as in VLBI
  monitoring by the MOJAVE program.  The dashed line shows where these
  two measures are equal.}
\label{betaappfig}
\vspace{2.2mm}
\end{figure}

Once $\dD$ and $\theta$ are determined from my method, the apparent
speed $\beta_{\rm app}$ one expects can be computed (Equation
[\ref{betaappeqn}]).  This can then be compared with the speeds of
knots seen in VLBI jet monitoring programs.  The MOJAVE program
regularly monitors a number of blazars with the VLBA.  On their
website they provide the median of the knot apparent speeds when jet
speeds for at least 5 knots can be computed.  There are 34 objects in
my sample that meet this criterion.  In Figure \ref{betaappfig} I
compare my results for $\beta_{\rm app}$ with the median apparent jet
speeds from the MOJAVE program.  The results are consistent for most
sources with in the errors, although my errors are large.  For some
sources, however, agreement is quite poor.  The blazar with the
fastest knot in the MOJAVE sample, PKS 0805$-$07 \citep{lister16}
actually has a fairly slow core speed based from my determination
($\beta_{\rm app} = 2.6^{+4.4}_{-1.7}$), and the median MOJAVE speed
is not consistent with my result.

\begin{figure}
\vspace{10.0mm} 
\epsscale{1.0} 
\plotone{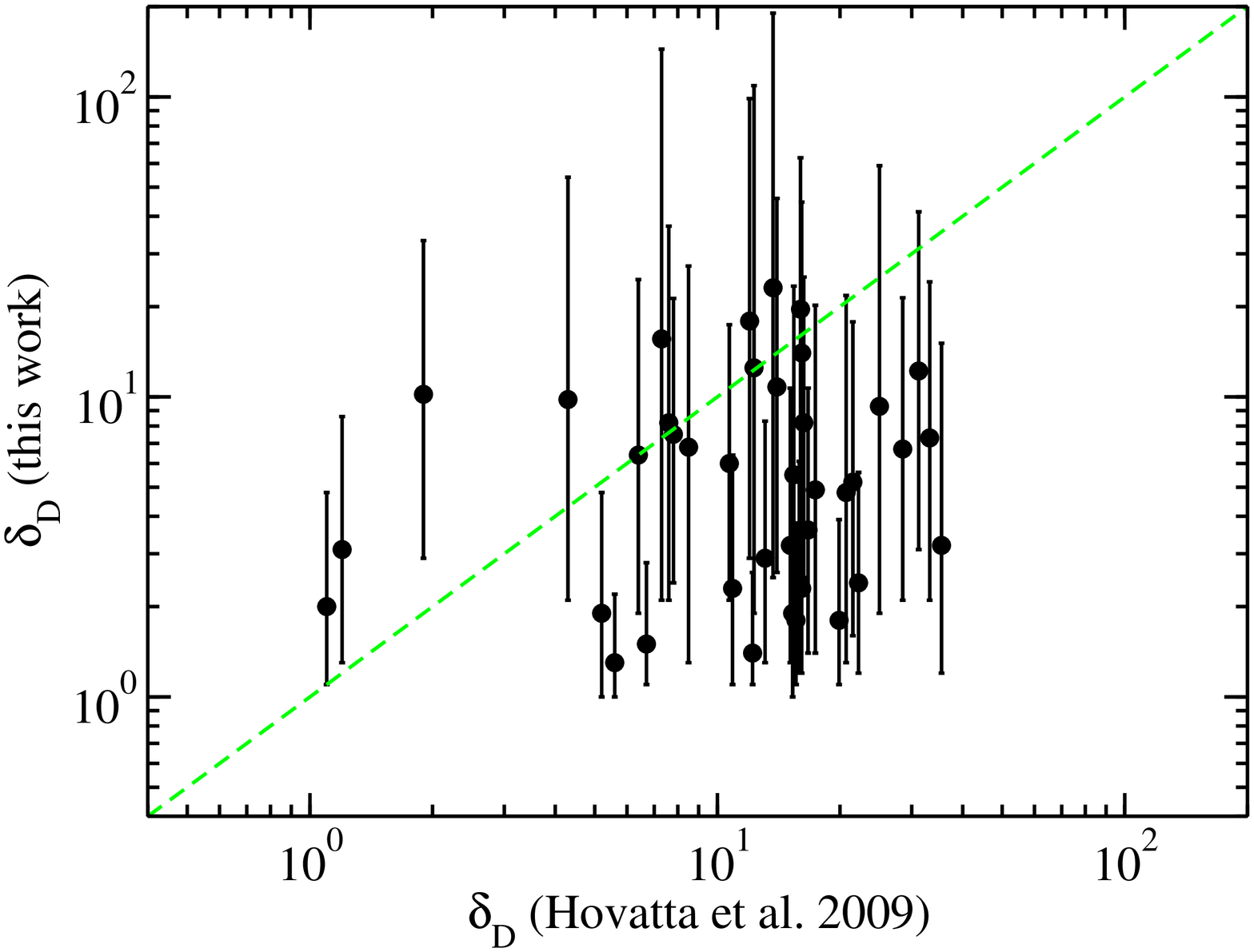}
\caption{Variability Doppler factors from \citet{hovatta09} compared
  with this work.  Dashed line is where the two methods for determining
  Doppler factor would be equal.}
\label{hovattadoppler}
\end{figure}

In Figure \ref{hovattadoppler} I compare my Doppler factor $\dD$
measurements to the variability Doppler factors measured by
\citet{hovatta09}.  Their Doppler factors are computed using radio
variability and brightness to determine an observed brightness
temperature.  This brightness temperature is compared to what one
would expect for the maximum intrinsic (unbeamed) brightness
temperature if it is limited by equipartition
\citep{readhead94}.  I plot my Doppler factors against those of
\citet{hovatta09} in Figure \ref{hovattadoppler}.  Again, for most
sources agreement is within the rather large errors.  However, there
is some evidence that equipartition may be violated during flares
\citep{homan06}.  Further, the components that flare may have
different Doppler factors than the core; knots can accelerate or
decelerate \citep[e.g.,][]{homan15,jorstad17}.

\begin{figure}
\vspace{10.0mm} \epsscale{1.0} \plotone{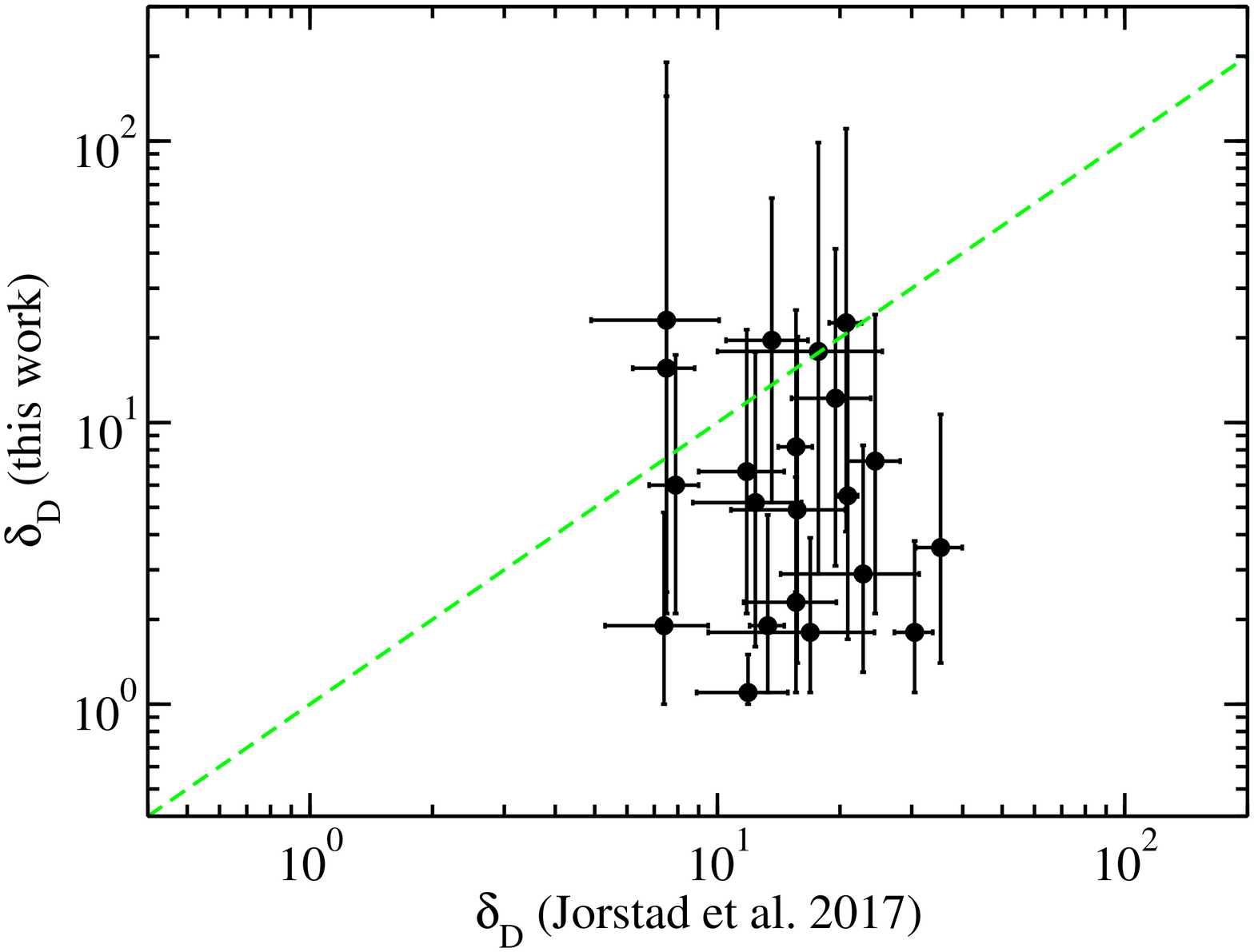}
\caption{Doppler factors from \citet{jorstad17} compared
  with this work.  Dashed line is the two ways of determining the Doppler
  factor would be equal.}
\label{jorstaddoppler}
\end{figure}

A different method for determining $\G$, $\theta$, and $\dD$ for
blazar jets was used by \citet{jorstad05,jorstad17}.  They routinely
monitor a number of blazars at 43 GHz with VLBA, and use kinematics of
observed knots to determine $\beta_{\rm app}$.  The Doppler factors of the
individual knots are determined by measuring the timescale for the
flux variations and assuming this variability timescale is limited by
the size of the knot, which can also be measured from the VLBA images.
Once they measure $\beta_{\rm app}$ and $\dD$ for a knot, they can
compute $\theta$ and $\Gamma$.  They computed the average jet
parameters for all the knots for each source.  In Figure
\ref{jorstaddoppler} I plot the Doppler factors from my calculation
versus the average Doppler factors determined by \citet{jorstad17} for
the sources where our samples overlap.  The agreement is clearly quite
poor.  This could be due to acceleration or deceleration of jet
components, or other sorts of variability.  The method of determining
$\dD$ from variability used by \citet{jorstad05,jorstad17} assumes the
variability timescale is dominated by the light-crossing timescale,
which might not be the case.

\begin{figure}
\vspace{10.0mm} 
\epsscale{1.0} 
\plotone{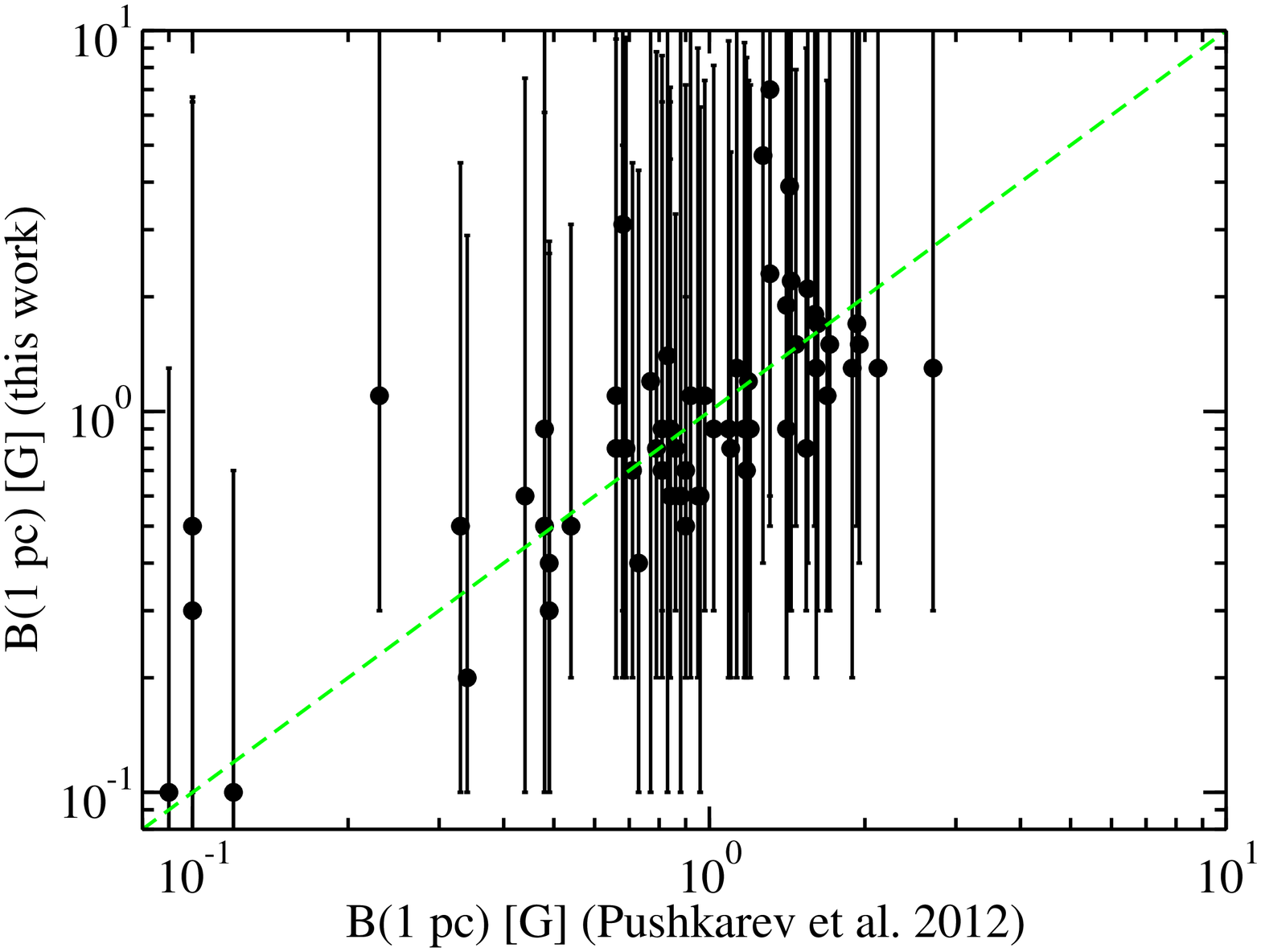}
\caption{The magnetic field at 1 pc from this work plotted against the
  same quantity as measured by \citet{pushkarev12}.  The dashed line
  shows where these two measures are equal.}
\label{Bmagfig}
\end{figure}

Using the core shift measurements, and assuming the BK model with
$\dD=\G$ and using the maximum jet speeds from the MOJAVE program,
\citet{pushkarev12} estimate the magnetic field strength at 1 pc.
They also make the assumption of equipartition between electrons and
magnetic field ($\xi_e=1$ in my notation).  In Figure \ref{Bmagfig} I
compare my magnetic field values with theirs.  My results are
consistent, within the errors, for all sources except one
(1334$-$127).  This is perhaps not surprising, considering both my
calculation and the one of \citet{pushkarev12} use the same core shift
data, although we make different assumptions.  The magnetic field
values may pose problems for modeling the multiwavelength SEDs of
blazars \citep{nalewajko14}.

\subsection{Proxies for Jet Angle}
\label{jetanglesection}

\begin{deluxetable*}{lccccccc}
\tabletypesize{\scriptsize}
\tablecaption{Correlations results}
\tablewidth{0pt}
\tablehead{
\colhead{y} & 
\colhead{x} & 
\colhead{Spearman} & 
\colhead{Kendall} & 
\colhead{F-test} & 
\colhead{$m$} & 
\colhead{$y_n$} & 
\colhead{$x_{n}$}
}
\startdata
$\log_{10}\G$ & $\log_{10}L_\g$ &   $1.8\sigma$ &   $1.9\sigma$ &   $5.4\sigma$ &   $0.12\pm0.03$ &   $0.21\pm0.05$ & 54 \\
$\log_{10}\theta$ &  $\log_{10}L_\g$ &  $2.8\sigma$ &   $2.8\sigma$ &   $2.9\sigma$ &  $-0.18\pm0.08$ &   $1.07\pm0.11$ & 54 \\
$\log_{10}\alpha$ &  $\log_{10}L_\g$ &   $2.3\sigma$ &   $2.4\sigma$ &   $2.6\sigma$ &   $-0.14\pm0.07$ &   $0.19\pm0.09$ & 54 \\
$\log_{10}B(1\ \pc)$ & $\log_{10}L_\g$ &  $6.3\sigma$ & $>8.3\sigma$ &  $7.5\sigma$ &  $0.27\pm0.14$ &   $-0.39\pm0.21$ & 54 \\
$\log_{10}\dD$ & $\log_{10}L_\g$ & $2.5\sigma$ & $2.5\sigma$ &  $5.4\sigma$ &  $0.13\pm0.03$ &   $0.27\pm0.04$ & 54 \\
$\log_{10}\beta_{\rm app}$ & $\log_{10}L_\g$ &   $1.1\sigma$ &   $1.1\sigma$ &   $1.6\sigma$ &  $0.12\pm0.11$ &  $-0.01\pm0.17$ & 54 \\
\hline
$\log_{10}\theta$ & $\log_{10}CD$ &  $4.5\sigma$ &  $4.3\sigma$ &  $5.0\sigma$ &  $0.67\pm0.11$ &  $-0.41\pm0.18$ & 2.0\\
$\log_{10}\theta$ & $\Delta\phi$ &  $6.9\sigma$ &   $>8.3\sigma$ &   $6.4\sigma$ &   $0.32\pm0.14$ &   $4.6\pm1.6$  & 0.10 \\
$\log_{10}\Phi_{\rm jet}/M_{\rm BH}$ & $\log_{10}L_{\rm acc}$ &   $2.7\sigma$ &   $2.7\sigma$ &   $4.3\sigma$ &   $0.28\pm0.16$ &   $33.01\pm0.28$ & 47.1 \\
$\log_{10}\theta$ & $\log_{10}\nu_{\rm pk}$\tablenotemark{a} &  $1.5\sigma$ &  $1.5\sigma$ &  $1.6\sigma$ &   $0.36\pm0.35$ &   $0.77\pm0.29$ & 14 \\
$\log_{10}\theta$ & $\log_{10}\nu_{\rm pk}$\tablenotemark{b} &  $0.9\sigma$ &  $1.2\sigma$ &   $1.6\sigma$ &   $0.27\pm0.27$ &   $0.69\pm0.32$ & 14 \\
\enddata
\tablenotetext{a}{\citet{meyer11} BL Lacs}
\tablenotetext{b}{3LAC BL Lacs}
\label{table_correlate}
\end{deluxetable*}

\begin{figure}
\vspace{10.0mm} 
\epsscale{1.0} 
\plotone{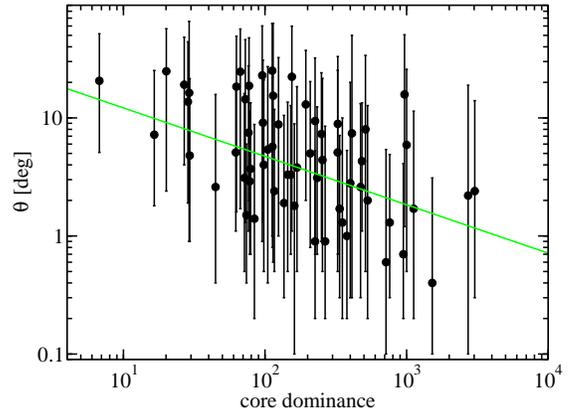}
\caption{The result $\theta$ determined here plotted versus CD.  The
  line shows the best fit.}
\label{thetaCD}
\vspace{2.2mm}
\end{figure}

The core dominance (CD)--the ratio of the core to extended radio
luminosity--has been used as a proxy for $\theta$
\citep[e.g.,][]{orr82,meyer11,marin16}.  I define CD as the ratio of
the core luminosity at 15 GHz (as reported by MOJAVE) to the extended
radio luminosity at 300 MHz \citep[as reported by][]{meyer11}.  The CD
as a function of the $\theta$ determined here is plotted in Figure
\ref{thetaCD}.  I test whether CD is correlated with $\theta$ and the
results are in Table \ref{table_correlate}.  In all cases the
significance is $<5\sigma$.  Again, I note that my errors on
$\theta$ are quite large.

One might also expect the core shift ($\Delta\phi$) to be correlated
with $\theta$.  I explore this correlation in Figure \ref{thetadphi}
and Table {\ref{table_correlate}.  The viewing angle $\theta$ is much
  more strongly correlated with $\Delta\phi$ ($>5\sigma$ for all
  tests) than CD.  I conclude that $\Delta\phi$ is a better proxy for
  $\theta$ than CD.  For sources where $\Delta\phi$ is measured, but
  other measurements needed to use my method are not, my resulting
  linear fit might be a useful way to estimate $\theta$.

\begin{figure}
\vspace{10.0mm} 
\epsscale{1.0} 
\plotone{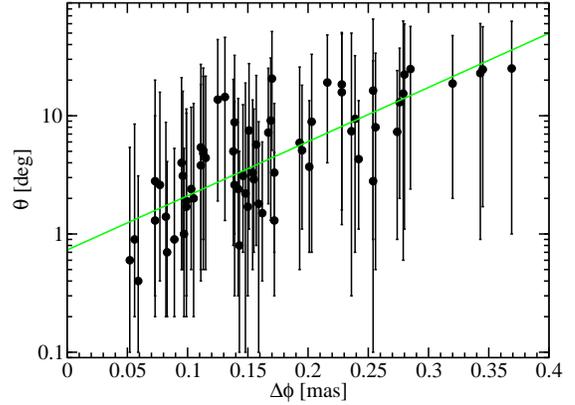}
\caption{The result $\theta$ determined here plotted versus
  $\Delta\phi$.  The line shows the best fit.}
\label{thetadphi}
\vspace{2.2mm}
\end{figure}

\subsection{Gamma Rays}
\label{gammaraysection}

TeV-detected BL Lac objects are often found to have knots moving at
low $\beta_{\rm app}$\ indicating low $\G$ and $\dD$
\citep{marscher99,piner04,piner05,piner08,piner10}.  This is in
contrast to multiwavelength SED modeling of these sources, which finds
much larger values of $\G$ and $\dD$
\citep[e.g.,][]{finke08_SSC,abdo11_mrk421,abdo11_mrk501,inoue16}.
This discrepancy is sometimes called the ``TeV Doppler factor
crisis''.  For almost every source in my sample $\G\la4$ is within the
quite large 68\% confidence interval.  Most notably, the two nearest
BL Lac objects, Mrk 421 and Mrk 501 have well-constrained low $\G$ and
$\dD$, and sub-luminal implied $\beta_{\rm app}$.  Several possible
resolutions to the TeV Doppler factor crisis have been suggested: the
speed of the jet could be stratified, with a slower layer to explain
the low speed from the radio, and a faster spine to explain the
multiwavelength emission \citep{ghisellini05}.  Or the jet could be
decelerating, with the faster part closer to the jet explaining the
multiwavelength emission and the slower part farther from the jet
explaining the radio emission \citep{georgan03}.  Finally, the overall
flow could have a low $\G$ consistent with radio observations, but
magnetic reconnection could lead to the creation of a ``jet within a
jet'' with large $\G$ to explain the multiwavelength emission
\citep{giannios09}.  Mrk 421 and Mrk 501 have values consistent with
large angles, but also are consistent with relatively small angles to
the line of sight ($\theta>9.9\arcdeg$ and $\theta>4.2\arcdeg$,
respectively, at 68\% confidence).  Large $\theta$ favors the jet
within a jet scenario, since in this model the overall jet could be
misaligned, but the jet within a jet could be oriented towards the
observer.  The other explanations require $\theta$ to be small.  The
small sample and large errors prevent me from making definitive
conclusions.

Many blazars are constrained to have low $\dD$ (Table
\ref{tableresult}) and are not detected at TeV energies (e.g., PKS
0607$-$15, PKS 1936$-$15).  If low $\dD$ is an indication of
brightness at very high energies, these sources could be potential TeV
sources, and observation of them with atmospheric Cherenkov telescopes
could result in detections.  However, many of them are at high
redshifts so absorption by the extragalactic background light
\citep[e.g.,][]{finke10_EBLmodel} could make them undetectable.

Although relatively few blazars have been detected with atmospheric
Cherenkov telescopes, a much larger number have been detected by the
\fermi\ Large Area Telescope (LAT).  Indeed, 54 of the 64 sources in
my sample are in the Third LAT AGN Catalog
\citep[3LAC;][]{ackermann15_3lac}.  I test correlations between the
$\g$-ray luminosity from this catalog with all of the parameters
presented in Table \ref{tableresult}.  The results can be seen in
Table \ref{table_correlate}.  The strongest correlation is found
between $B(1\ \pc)$ and $L_\g$.  This result is plotted in Figure
\ref{LgBmagplot}.  This is perhaps not a surprise. The magnetic field
$B$ is correlated with $P_j$ (Equation [\ref{Pj2}]), which is in turn
determined from $L_{\rm ext}$ (Section \ref{jetpowersection}).  The
respective luminosities $L_{\rm ext}$ and $L_\g$ could be correlated
due to both depending on distance.  The parameters $\theta$ and $\dD$
also show strong correlations with the F-test, but not the
non-parametric tests.

\begin{figure}
\vspace{10.0mm} 
\epsscale{1.0} 
\plotone{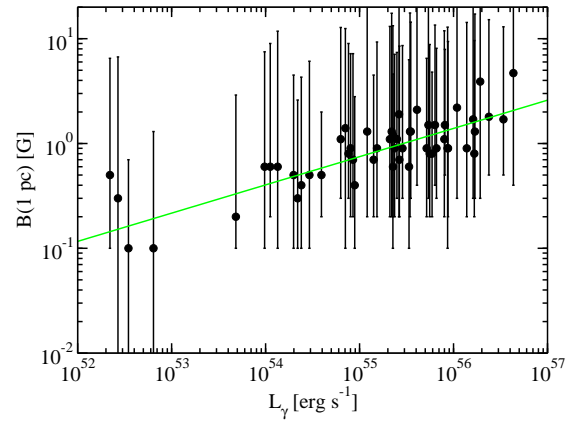}
\caption{A plot of $B(1\ \pc)$ versus LAT $\g$-ray luminosity
from the 3LAC.   The line shows the best fit.}
\label{LgBmagplot}
\vspace{2.2mm}
\end{figure}

\subsection{Implications for Jet Physics}
\label{jetphysics}

General relativistic magnetohydrodynamic simulations of jets launched
from magnetically arrested disks (MADs) indicate that energy can be
extracted from the rotation of black holes to form jets by the
Blandford-Znajek mechanism \citep{blandford77} that appear very
similar to the ones found in nature \citep{tchek11}.  The parsec-scale
magnetic flux $\Phi_{\rm jet}$ can be determined from the jet parameters I have
computed by
\begin{flalign}
\label{phijet1}
\Phi_{\rm jet} = 1.2\times10^{34}\ \G\alpha\left[\frac{M_{\rm BH}}{10^9\ M_\odot}\right]
\left[\frac{B(1\ \pc)}{1\ \Gauss}\right]\ \Gauss\ \cm^2
\end{flalign}
where $M_{\rm BH}$ is the black hole mass, making
  assumptions about equipartition \citep{zam14}.  A more detailed
  calculation by \citet{zdz15} gives the more general
\begin{eqnarray}
\label{phijet1accurate}
\Phi_{\rm jet} & = 8.2\times10^{33}\ 
\G\alpha\left[\frac{M_{\rm BH}}{10^9\ M_\odot}\right]
\left[\frac{B(1\ \pc)}{1\ \Gauss}\right]\ 
\nonumber \\ & \times
\left[\frac{g_B\hat\g_{\rm ad} + \xi_e + \xi_p + \hat\g_{\rm ad}\xi_m}
{\xi_e + \xi_p + \hat\g_{\rm ad}\xi_m}\right]^{1/2}
\Gauss\ \cm^2
\end{eqnarray}
where I have rewritten their result using Equation
(\ref{magnetization}).  The theory of jets launched from MADs
predicts a relationship between $\Phi_{\rm jet}$ and the accretion
disk luminosity $L_{\rm acc}$,
\begin{flalign}
\label{phijet2}
\Phi_{\rm jet} & = 2.4\times10^{34} \left[\frac{M_{\rm BH}}{10^9\ M_\odot}\right]
\nonumber \\ & \times
\left[\frac{L_{\rm acc}}{1.26\times10^{47}\ \erg\ \s^{-1}}\right]^{1/2}\ \Gauss\ \cm^2
\end{flalign}
\citep{zam14}.  In Figure \ref{phijetfig} I plot $\Phi_{\rm
  jet}/M_{\rm BH}$ determined from Equation (\ref{phijet1accurate})
using jet parameters from this work versus $L_{\rm acc}$ as found by
\citet{zam14}.  I tested for the significance of the correlation of
these quantities (Table \ref{table_correlate}) and found weak
significance with the F-test, and no significance with the Spearman
and Kendall tests.  As Figure \ref{phijetfig} demonstrates, the best
fit line does not agree with the model prediction for jets launched
from MAD disks from \citet{zam14}, which overestimates my results by a
factor of $\approx 10$.  I have computed $\Phi_{\rm jet}$ for the
sources in my sample using both Equation (\ref{phijet1}) and
(\ref{phijet1accurate}), and the results are not significantly
different.  \citet{pjanka17} compared several estimates of jet power,
and found that computing it from extended radio luminosity gives a
factor of $\approx 10$ lower result than from core shift measurements
or from broadband SED modeling.  Since $B_0\propto \sqrt{P_{\rm jet}}$
(Equation [\ref{Pj2}]), when I scale up the jet powers by a factor of
10, I find that $\Phi_{\rm jet}$ for my sources increases by a factor
of $\approx\sqrt{10}\approx 3$.  This improves agreement with Equation
(\ref{phijet2}), but the computed results still do not agree with this
theoretical curve. The reason for the disagreement is not entirely
clear.  Equation (\ref{phijet1accurate}) assumes that all the black
holes have spin $a\approx1$ and Equation (\ref{phijet2}) assumes all
accretion disks have an accretion efficiency $\eta=0.4$, neither of
which may be the case \citep{zam14,zdz15}.  Another assumption may not
be correct, or these sources may not be accreting in the MAD regime.

\begin{figure}
\vspace{10.0mm} 
\epsscale{1.0} 
\plotone{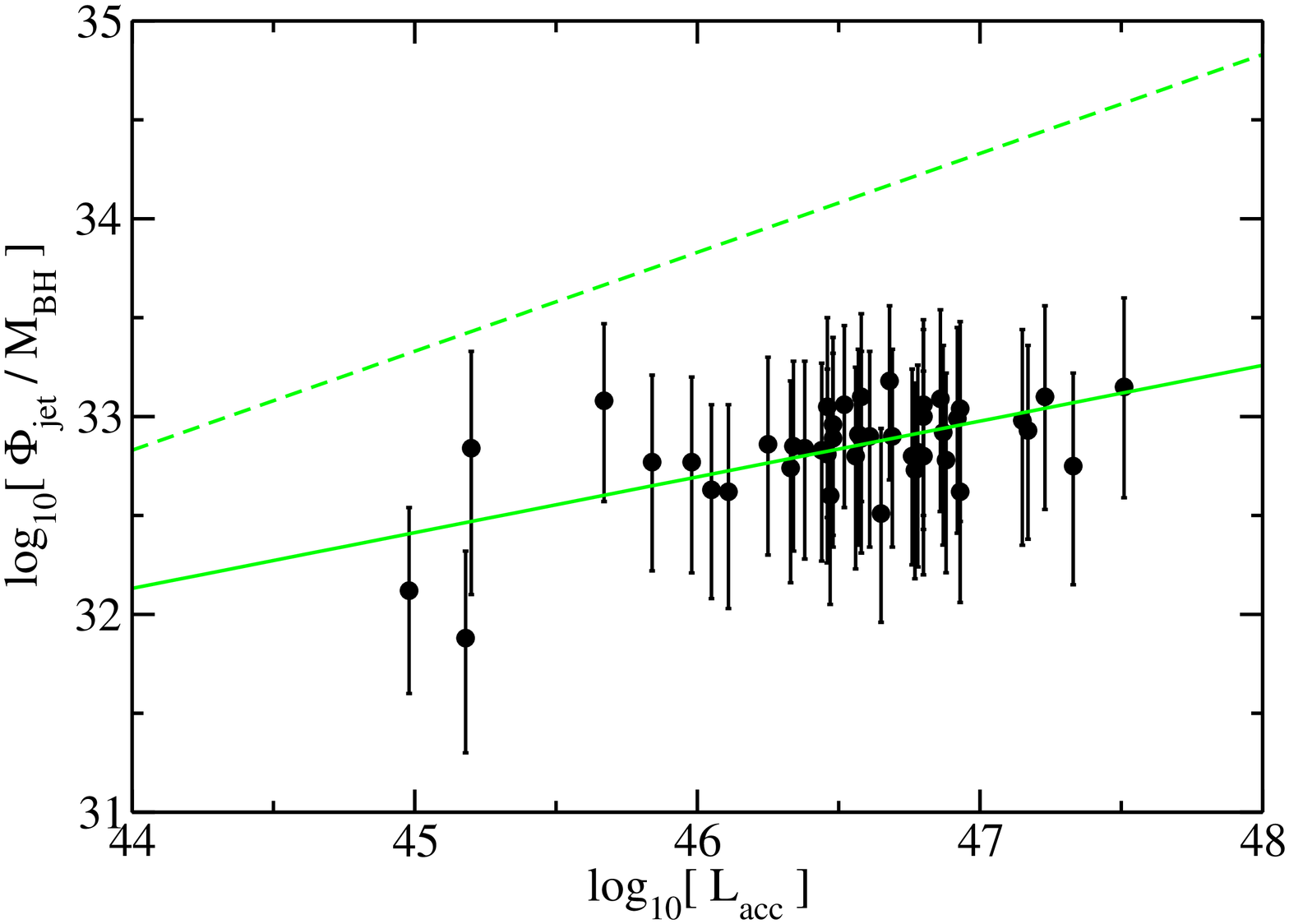}
\caption{Correlation between magnetic flux divided by black hole mass
  ($\Phi_{\rm jet}/M_{\rm BH}$) versus $L_{\rm acc}$.  The theoretical
  expectation from jets launched from a \citet{shakura73} accretion
  disk around a maximally rotating black hole (Equation
  [\ref{phijet2}]) is shown as the dashed line while the line from my
  best fit is shown as the solid line.  The parameter $\Phi_{\rm jet}$
  is in units $\Gauss\ \cm^{2}$, $M_{\rm BH}$ is in units
  $10^9M_\odot$, and $L_{\rm acc}$ is in units $\erg\ \s^{-1}$.}
\label{phijetfig}
\vspace{2.2mm}
\end{figure}

\citet{meyer11} have introduced a scenario where FSRQs have jets which
essentially have the same $\G$ for the whole jet length, and BL Lac
objects have decelerating jets.  For BL Lac objects, as $\theta$
increases, one sees slower parts of the jets with larger beaming
cones.  Their scenario explains the discrepancy between low Doppler
factors found in multiwavelength SED modeling of FR I radio galaxies
and the high Doppler factors found in modeling the multi-wavelength
SEDs of BL Lac objects \citep[e.g.,][]{chiaberge00}.  Their scenario
predicts that $\theta$ is correlated with the peak frequency
($\nu_{\rm pk}$) of the low-energy synchrotron component in the SEDs
of BL Lac objects.  In Figure \ref{thetanupk} I plot my determination
of $\theta$ versus $\nu_{\rm pk}$ with $\nu_{\rm pk}$ taken from
\citet{meyer11} and the Third LAT AGN Catalog
\citep[3LAC;][]{ackermann15_3lac}.  The correlation between $\theta$
and $\nu_{\rm pk}$ is not significant in any of my tests (Table
\ref{table_correlate}).  However, I note that the error on $\theta$
is quite large, and there are only 11 BL Lacs in my sample, and only
2 with $\log_{10}[\nu_{\rm pk}/\Hz]>14.5$.  This test is clearly not
definitive.

\begin{figure*}
\vspace{10.0mm} 
\epsscale{1.0} 
\plottwo{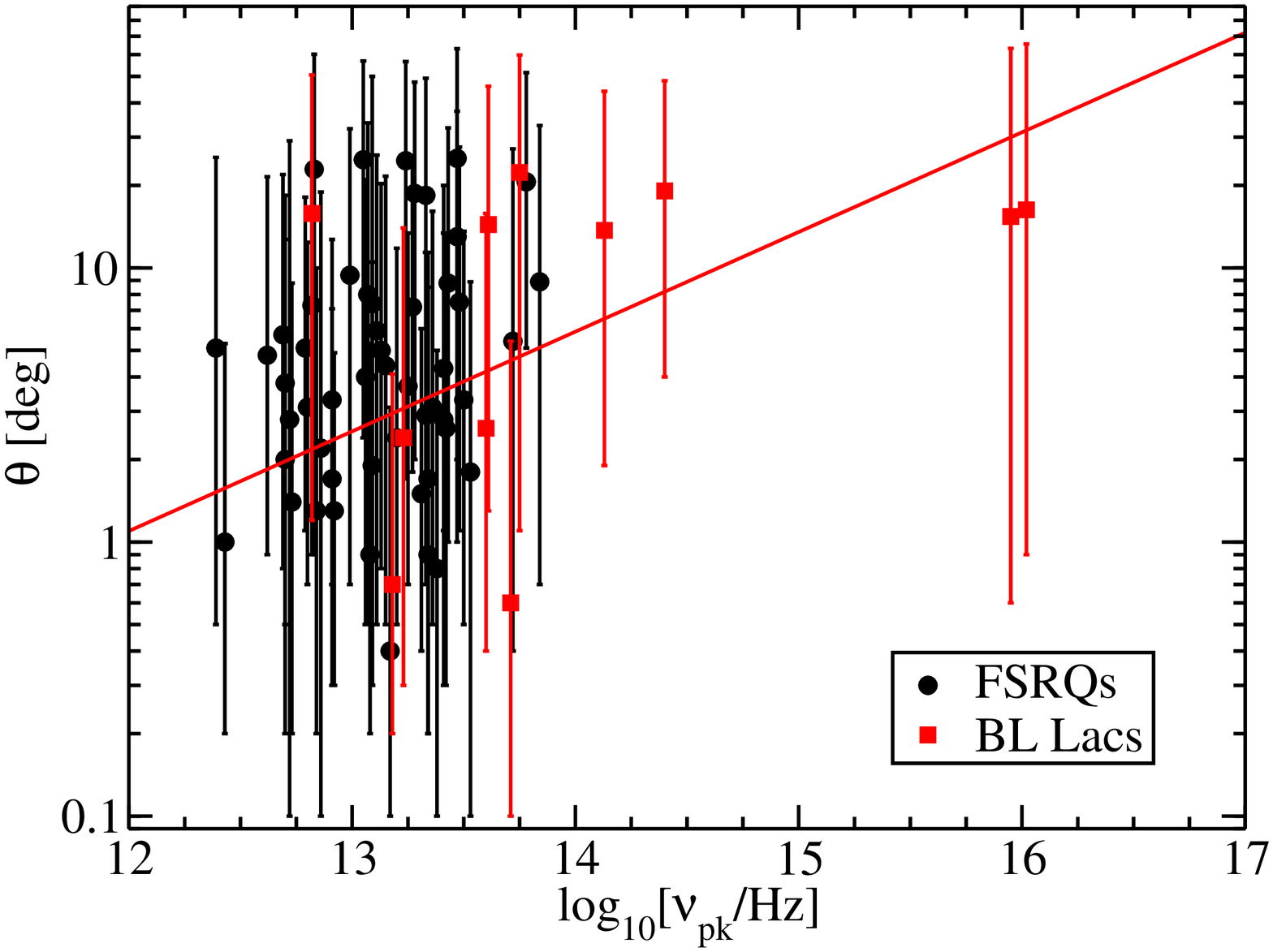}{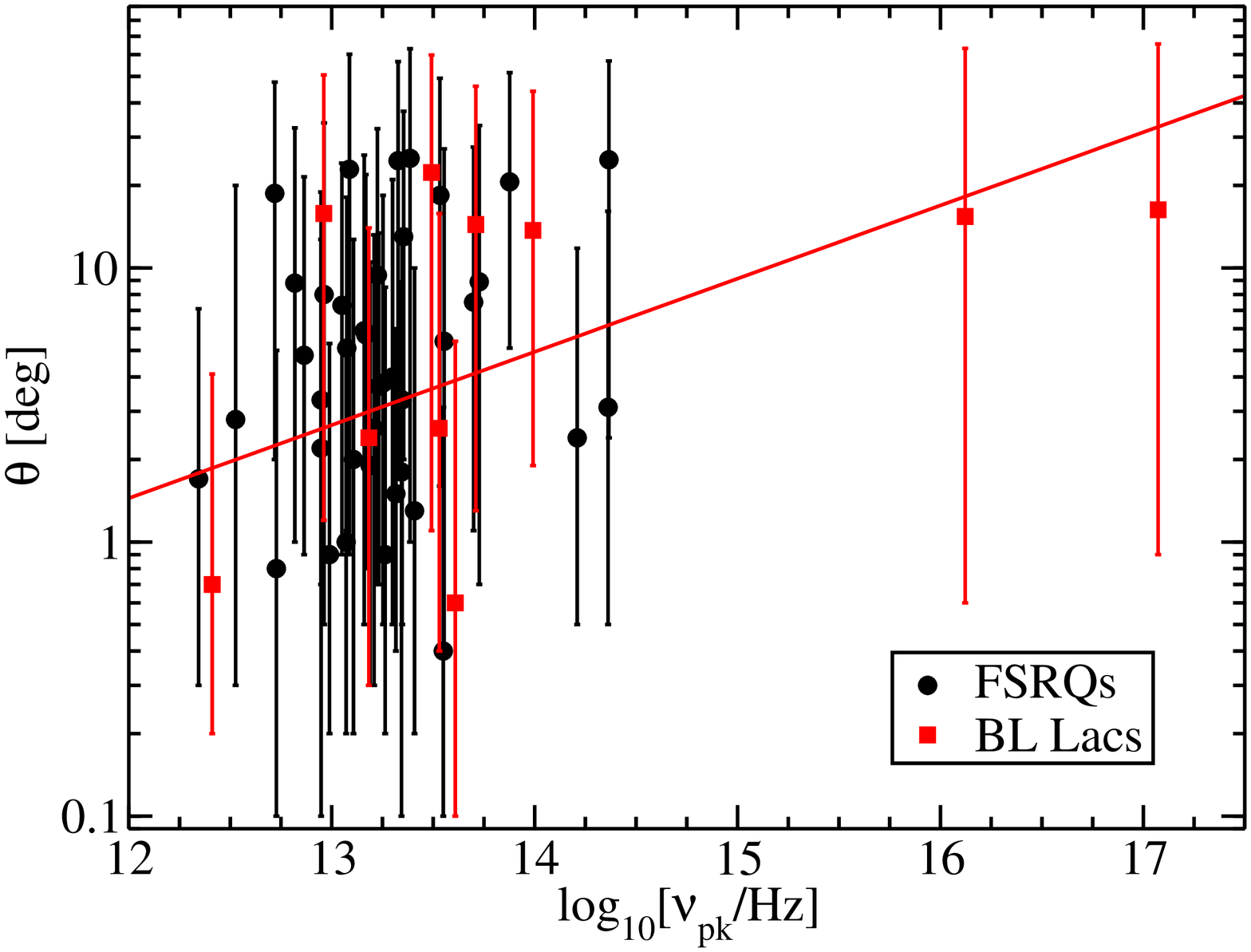}
\caption{My determination of $\theta$ versus $\log_{10}\nu_{pk}$ for FSRQs
  (black circles) and BL Lac objects (red squares).  The line
  indicates the best fit for BL Lac objects.  Left:
  $\log_{10}\nu_{\rm pk}$ from \citet{meyer11}.  Right:
  $\log_{10}\nu_{\rm pk}$ from 3LAC \citep{ackermann15_3lac}.}
\label{thetanupk}
\vspace{2.2mm}
\end{figure*}

It is expected in the BK model that there is an inverse relationship
between the jet Lorentz factor and opening angle, i.e., that
$\alpha\G$ is a constant for all sources in the BK jet model.  This
parameter is important for a number of processes in jet physics
\citep[see][and references therein]{clausen13}.  A constant $\alpha\G$
has been found by \citet{jorstad05,jorstad17},
\citet{pushkarev09,pushkarev17}, and \citet{clausen13}.  I plot
$\alpha\G$ for all the sources in my sample in Figure
\ref{gammaalphafig}.  I perform a fit to $\alpha\G=\zeta$ instead of
plotting and fitting $\alpha = \zeta/\Gamma$ in order to take into
account the correlation in the errors on $\alpha$ and $\Gamma$.  I
find $\zeta = 0.054\pm0.014$ \ with $\chi^2$/dof = $4.1/63$ ,
certainly consistent with a constant $\alpha\G$.  This value is lower
than typically found by other authors.  \citet{jorstad05} found
$\zeta=0.17\pm0.08$ for their sample, and more recently
\citet{jorstad17} found $\zeta=0.19\pm0.07$ and $\zeta=0.32\pm0.13$
for two different ways of determining $\alpha$.  \citet{pushkarev09}
and \citet{pushkarev17} found median $\zeta=0.13$ and $\zeta=0.175$,
respectively, in their samples; \citet{clausen13} found
$\zeta\approx0.2$ from their sample.

It is thought that $(\G\alpha)^2<\sigma$
\citep[e.g.,][]{tchek09,zdz15,pjanka17}.  Since, based on my priors,
the magnetization parameter, $\sigma$ (Equation [\ref{magnetization}])
ranges from $0.006$ to $200$, and I find $(\Gamma\alpha)^2 = \zeta^2 =
0.0029\pm0.0015$, my results indicate that indeed
$(\G\alpha)^2<\sigma$.

\begin{figure}
\vspace{10.0mm} \epsscale{1.0} \plotone{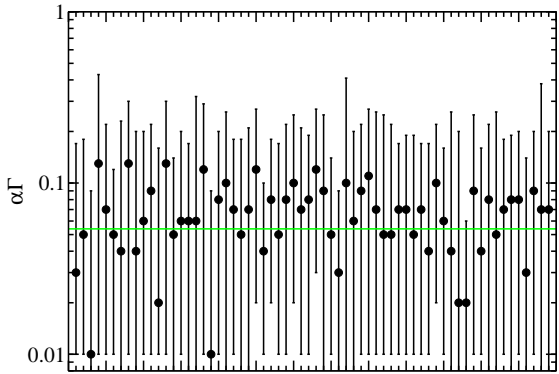}
\caption{The value $\G\alpha$ for all sources in my sample.  The line
shows the best fit value.}
\label{gammaalphafig}
\vspace{2.2mm}
\end{figure}

\section{Discussion}
\label{discussion}

I have shown that using five observables ($z$, $F_\nu$, $\Delta\phi$,
$\alpha_{\rm app}$, $L_{\rm ext}$) with the BK model, it is possible
to determine $\theta$ and $\G$ and other properties for parsec-scale
blazar jets.  These results are generally consistent with other
constraints on $\theta$, $\G$, and $\dD$, although my errors are quite
large.  This limits my method's usefulness.  With some exploration, I
find that my uncertainties are dominated by two sources:
\begin{itemize}
\item The errors on the core shift measurement ($\Delta\phi$) are
  large, $\approx 15-50\%$.  These could be improved by measuring core
  shifts at multiple frequencies, and doing a fit to these data.
  \citet{sokolovsky11} have done this, although they measure the core
  shifts with a different technique, and have a much smaller sample
  size than \citet{pushkarev12}.  Also, there is the issue of
  validating $\Delta\phi$ measured with different techniques.  For
  instance, for 2201+315, the fit to core shift measurements at 6
  frequencies from \citet{sokolovsky11} results in
  $\Delta\phi=0.188\pm0.009$\ masec between 15 and 8 GHz, while
  \citet{pushkarev12} measure a discrepant
  $\Delta\phi=0.345\pm0.051$\ masec.
\item The uncertainty in the electron spectral index ($p$), which I
  draw from a flat prior.  This could in principle be measured from
  the SEDs of a blazars.  However, practically, it is unclear if one
  could distinguish the parsec-scale portion of the jet from other,
  more compact, highly variable components that dominate the SED of
  blazars at high frequencies.  Alternatively, one could compute $p$
  from $\G$ using shock physics and results from test-particle
  relativistic shock acceleration theory \citep{keshet05}.  However,
  this may not be applicable to realistic shocks, where
  nonlinear effects could be important.  I performed calculations with
  $p$ constrained by the formula of \citet{keshet05}, and found the
  resulting $\theta$ for some sources to be unrealistically large.
  For example, for 1101+384 (Mrk 421) I found
  $\theta=60^{+10}_{-12}$\ deg, inconsistent with the jet/counter-jet
  brightness ratio constraint for this source \citep{piner05}, and the
  general expectation that blazars have small $\theta$.

\end{itemize}

It has also been questioned how reliable it is to use the extended
radio luminosity as a proxy for jet power \citep[e.g.,][]{godfrey16}.
However, I found that this is not likely to be a major source of
error, at least compared to uncertainties on $\Delta\phi$ and $p$ (see
Section \ref{altjetpower}).  \citet{pjanka17} compared
  several methods of estimating jet power: from extended radio
  luminosity (the method used here), from core shift measurements, and
  based on broad-band SED modeling.
  Each technique has its own set of assumptions.  Since authors rarely
  provide error estimates on jet powers, it is difficult to compare
  these methods; however, \citet{pjanka17} found that the core shift
  and SED modeling jet powers agreed on average, and that these
  methods generally gave values $\approx10$ larger than the extended
  radio luminosity method.  Besides issues with model assumptions, the
  discrepancy could be due to short term power measured with SED
  fitting and core shifts, versus long-term power measured with the
  lobes; or the core shift and SED fitting powers could be lower due
  to having more electron/positron pairs relative to protons than
  assumed in these methods \citep[see also][]{inoue17}.  

Aside from these uncertainties, there is also the problem of
variability.  At higher frequencies blazars are extremely variable,
often with fluxes varying by several orders of magnitude.  At radio
frequencies, they are less variable, but their fluxes can still vary
by $\sim$\ a few.  The core shifts could also vary with time.  I have
used measurements of core fluxes and core shifts that are
simultaneous.  However, since the BK model is an approximation for a
variable jet, with a number of colliding shells, there is another
source of error associated with the limitations of this model.

Despite these issues I do think this method can be a useful way
to constrain jet parameters, complementary to other methods.  This
will be particularly true if ways to mitigate the uncertainties
discussed above can be found.

\acknowledgements 

I thank the referee for valuable comments that have improved this
manuscript.  I am grateful to Matthew Lister for several useful
  discussions about determining jet parameters, and Tuomas Savolainen
  for a useful discussion about the calculation of magnetic flux.
This research has made use of data from the MOJAVE database that is
maintained by the MOJAVE team \citep{lister09}.  I am supported by the
Chief of Naval Research and NASA under contract S-15633Y.


\bibliographystyle{apj}
\bibliography{variability_ref,EBL_ref,references,mypapers_ref,blazar_ref,sequence_ref,SSC_ref,LAT_ref,3c454.3_ref,radiolobe_ref}

\end{document}